\documentclass[12pt]{article}
\usepackage{hhline}
\usepackage{latexsym,graphicx}
\usepackage{subfigure}
\usepackage{amssymb}
\usepackage{amsmath}
\usepackage{amscd}
\usepackage{amsthm}
\usepackage{float}
\usepackage[left=2cm,top=2.5cm,right=2.5cm,bottom=1.5cm]{geometry}   
 
\usepackage{xcolor}
\begin{document}
\begin{center}
\large{\bf{Note on Dark Energy and Cosmic Transit in a scale-invariance cosmology}} \\
\vspace{10mm}
\normalsize{Nasr Ahmed$^{1,2}$ and Tarek M. Kamel$^2$ }\\
\vspace{5mm}
\small{\footnotesize $^1$ Mathematics Department, Faculty of Science, Taibah University, Saudi Arabia } \\
\small{\footnotesize $^2$ Astronomy Department, National Research Institute of Astronomy and Geophysics, Helwan, Cairo, Egypt\footnote{abualansar@gmail.com}} \\
\end{center}  
\date{}
\begin{abstract}

In general, the laws of physics are not invariant under a change of scale. To find out whether the 'scale-invariance hypothesis' corresponds to nature or not, a careful examination to its implications is required. As a consequence, the scale-invariance cosmological models need to be carefully checked with many tests in order to confirm or disconfirm them. In this paper, three different toy models have been introduced in the framework of a scale-invariance cosmology to examine dark energy and cosmic transit. Although cosmic transit exists in the three models, the pressure stays always negative during cosmic evolution. In addition, there is always a singularity in the evolution of the equation of state parameter which is not suitable for a complete investigation of dark energy evolution. The undesirable features of the parameters have been discussed, and a comparison with other cosmological contexts has been done.

\end{abstract}

\section{Introduction and motivation}

The discovery of the accelerated expansion of the universe has been considered as one of the most challenging problems in modern physics \cite{11,13,14}. In order to find a satisfactory explanation, several proposals have been suggested among them is the dark energy (DE). Dark energy is assumed to be an exotic form of energy with negative pressure which acts as a repulsive gravity and, consequently, pushes the the universe to expand faster. Some dynamical scalar field models for DE have been introduced including quintessence \cite{quint}, Chaplygin gas \cite{chap}, phantom energy \cite{phant}, k-essence \cite{ess}, tachyon \cite{tak}, holographic \cite{holog1, holog2} and ghost condensate \cite{ark,nass}. Another approach to explain this cosmic acceleration is by modifying the geometrical part of Einstein-Hilbert action \cite{moddd}. It has been shown that such 'modified gravity theories' can explain the galactic rotation curves without assuming the existence of dark energy \cite{noji, cap, de}. Examples of this approach include $f(R)$ gravity \cite{39} in which the Lagrangian is a function of the Ricci scalar $R$, Gauss-Bonnet gravity \cite{noj8}, $f(T)$ gravity \cite{torsion} where $T$ is the torsion scalar, and $f(R,T)$ gravity \cite{1}, where $T$ is the trace of the energy momentum tensor ( see \cite{add1} for a detailed review ). In $f(T)$ gravity, the torsion scalar $T$ is used in the Einstein-Hilbert action instead of the Ricci scalar $R$. An advantage of $f(T)$ gravity is that it has second-order field equations while $f(R)$ gravity has fourth-order equations. In Gauss-Bonnet gravity, the Ricci scalar $R$ in the action is replaced by the the Gauss-Bonnet term $G=R^2-4R^{\mu\nu}R_{\mu\nu}+R^{\mu\nu\rho\delta}R_{\mu\nu\rho\delta}$, i.e. a function $f(G)$ can be used. \par

The equation of state parameter $\omega$ is the division of cosmic pressure and energy density, $\omega=\frac{p}{\rho}$. Investigating the evolution of this parameter is very essential to understand the nature of dark energy. This parameter is equal to $0$ for dust, $1/3$ for radiation and $-1$ for vacuum energy (cosmological constant). It can be less than $-1$ in scalar field models such as phantom $\omega \leq -1$, quintessence $-1 \leq \omega \leq 1$ and quintom which can go behind the phantom divide line $\omega = -1$. A no-go theorem has been suggested in quintom cosmology \cite{nogo} forbids $\omega$ of a single perfect fluid in Friedmann-Robertson-Walker geometry to cross the $-1$ boundary. A simple quintom model can be constructed using two scalar fields with one being quintessence and the other being phantom. $\omega=1$ is the largest possible value consistent with causality, it is assumed to happen for some exotic type of matter called Zel’dovich fluid where the speed of sound is equal to the speed of light \cite{zeld}. A matter fluid with $\omega=\frac{p}{\rho} \gg 1$ can exist in Ekpyrotic cosmological models \cite{ko,yo}. Although recent observations supports $\omega_{\Lambda}=-1$, dynamical models of dark energy are still possible especially those with $\omega$ across $-1$ \cite{cai1}. \par

Other alternatives to dark energy have also been suggested \cite{alt1, alt3, basic1}. It has been shown in \cite{alt1} that a cosmological model consistent with observations can be obtained from a classical theory with massive gravitons without need to DE assumption. The alternative assumption of bulk viscosity has been proposed in \cite{alt3} where the effective pressure becomes negative for a sufficiently large bulk viscosity and could mimic a dark energy equation of state. \par

The idea of scale-invariance of the macroscopic empty space has been proposed and studied in \cite{basic1}. Since the laws of physics are not generally invariant under a change of scale, a careful examination to the implications of such 'scale-invariance hypothesis' is needed to find out if it corresponds to nature or not \cite{bass}. It has also been noted in \cite{basic1} that scale-invariance cosmological models need to be further carefully checked with many tests in order to confirm or disconfirm them, this was the main motivation behind the current work. In the current work, we test the scale-invariance cosmology suggested in \cite{basic1} through three toy models using three different empirical ansatze. The three ansatze have been proved to be consistent with observations in several publications as we will see. We then investigate the physical behavior of the energy density, cosmic pressure, and the equation of state parameter.  The work plan and main points of the this paper can be summerised as follows: 

\begin{itemize} 
\item The main aim of the current work is to test the viability of the recently suggested 'scale-invariance cosmology' \cite{basic1} by investigating the cosmic transit and dark energy assumption in its framework. 

\item In order to do this, we have introduced three models based on three different ansatze. These ansatze have been used before in several cosmological studies because of their consistency with observations, represented mainly in the evolution of both the deceleration and jerk parameters. They all lead to a deceleration-acceleration cosmic transit and a flat Lambda CDM universe at late time.

\item Dark energy is usually defined as an exotic form of energy with negative pressure. If the deceleration-acceleration cosmic transit exists in a certain model, the evolution of cosmic pressure in that model should reveal a positive-to-negative sign flipping in corresponding to the positive-to-negative sign flipping of the deceleration parameter. Also, the evolution of the EoS parameter shouldn’t reveal any improper behaviour or singularities.

\item	The behaviors of cosmic pressure $p$ and EoS parameter $\omega$ in the three models have been found to be disappointing. The pressure is always negative with cosmic time which doesn't help to explain the cosmic transit, and the EoS parameter always shows a singularity which means that it is not possible to get a complete description for the evolution of dark energy.

\item	Because of such incomplete and undesirable behaviors of $p$ and $\omega$ in the three models with three different empirical forms of the scale factor, we conclude that the recently suggested scale-invariance cosmology in \cite{basic1} might not be able to provide a complete description for dark energy and cosmic evolution.

\item	The three ansatze used in this work provide excellent behavior of both $p$ and $\omega$ in the framework of other cosmological contexts and different gravity theories. In the last section, we have mentioned some examples of such theories where the same ansatze lead to a positive-to-negative sign flipping in cosmic pressure, and a complete evolution of EoS parameter with no singularities.
\end{itemize} 

The current work represents an attempt to test the scale-invariance cosmology through some ansatze which are shown to be observationally consistent and also have been successful in other cosmological contexts. Although the present work brings doubts to the viability of the scale-invariance cosmological models, more empirical forms of the scale factor still need to be utilized to make a final conclusion. Also, a future modified version of the scale-invariance cosmology may avoid the bad features we have got here. \par

The rest of the paper is organized as follows: In section 2, we give a review to the recently suggested scale-invariant gravity and the associated scale-invariance cosmology. In section 3, we investigate three different solutions to the spatially flat cosmological equations and analyze the behavior of different parameters. In section 4, we compare the obtained results with the results of other gravity theories where the same ansatze have been used. The final conclusion is included in section 5.

\section{scale-invariance gravity}

Scale-invariance means that the equations don't change under the line element transformation $ds^{'}=\lambda(t) ds$. where $ds^{'}$ is the line element of general relativity and $ds$ is the line element of a more general space \cite{basic01}. Although some modified scale-invariance gravity theories have been introduced to explain the cosmic acceleration \cite{basic0,basic1, basic}, the ability of the scale-invariance cosmological models to provide a complete description to cosmic evolution still not confirmed.\par The Robertson-Walker metric is written as
\begin{equation}
ds^{2}=dt^{2}-a^{2}(t)\left[ \frac{dr^{2}}{1-\kappa r^2}+r^2(d\theta^2+\sin^2\theta d\phi^2) \right] \label{RW}
\end{equation} 
where $r$, $\theta$, $\phi$ are comoving spatial coordinates, $a(t)$ is the cosmic scale factor, $t$ is time, $\kappa$ is either $0$, $-1$ or $+1$ for flat, open and closed universe respectively. In \cite{basic}, the conformal transformations 
\begin{equation} \label{conf}
\widetilde{g}_{\mu\nu}=e^Q g_{\mu\nu}
\end{equation}
have been applied to the Robertson-Walker metric where $Q$ is the quantum potential with the form
\begin{equation}
Q=\frac{3}{2} \left(q-\frac{1}{2}\right)\frac{H^2}{m^2}
\end{equation}
$q=\frac{\ddot{R}R}{\dot{R}^2}$ is the deceleration parameter, $H=\frac{\dot{R}}{R}$ is the Hubble parameter and $m$ is the particle's mass. The modified Einstein field equation now becomes \cite{basic0,basic1}
\begin{equation} \label{ein22}
\widetilde{R}_{\mu\nu}-\frac{1}{2}\widetilde{g}_{\mu\nu}\widetilde{R}=8\pi G \widetilde{T}_{\mu\nu}+\Lambda \widetilde{g}_{\mu\nu}
\end{equation}
where $\widetilde{R}_{\mu\nu}$ is the Ricci tensor with respect to the modified
metric $\widetilde{g}_{\mu\nu}$, $\widetilde{R}$ is the Ricci scalar, $\widetilde{T}_{\mu\nu}$ is the energy-momentum tensor, and $\Lambda$ is the effective cosmological constant. The energy-momentum tensor $\widetilde{T}_{\mu\nu}= \widetilde{T}^{(M)}_{\mu\nu}+ \widetilde{T}^{(Q)}_{\mu\nu}$ where $\widetilde{T}^{(M)}_{\mu\nu}$ is related to the matter contribution, and $\widetilde{T}^{(Q)}_{\mu\nu}$ arises from the energy density of the quantum potential. The solution of ($\ref{ein22}$) gives the following modified Friedmann equations
\begin{eqnarray}
\frac{\dot{R}^2}{R^2}&=&\frac{8\pi G}{3}\rho+\frac{\Lambda \lambda^2}{3}-2\frac{\dot{\lambda}\dot{R}}{\lambda R}+\frac{\dot{\lambda}^2}{\lambda^2}-\frac{\kappa}{R^2},  \\    \nonumber
\frac{\ddot{R}}{R}&=&-\frac{4\pi G}{3}(\rho+3p)+\frac{\Lambda \lambda^2}{3}-\frac{\dot{\lambda}\dot{R}}{\lambda R}-\frac{\dot{\lambda}^2}{\lambda^2}-\frac{\ddot{\lambda}}{\lambda}.
\end{eqnarray}
where $\lambda^2=e^Q$, for $\lambda=1$ we get the ordinary FRW equations. We consider only the flat case which is consistent with recent observations \cite{flatt}. While in \cite{basic} a universe with $\Lambda=0$ has been considered, we will use a very small positive value for the cosmological constant as suggested by observations.

\section{Cosmological solutions}

We are going to try three different solutions, each of them satisfies two observational conditions: 1- The deceleration-acceleration cosmic transit which means there is a sign flipping from positive to negative in the evolution of the deceleration parameter.\newline 
2- Since we are considering a flat universe ($\kappa =0$), the jerk parameter must have the asymptotic value $j=1$ at late-time. We recall that flat $\Lambda$CDM models have $j = 1$ \cite{81}. 

\subsection{The hyperbolic solution.}

The following ansatz gives a good agreement with observations for $0 < n < 1$
\begin{equation} \label{ansatz}
a(t)= A\sinh^n(b t)
\end{equation}
The main motivation behind using the ansatz (\ref{ansatz}) is its consistency with observations, and it has been used in constructing several cosmological models in different gravity theories \cite{pr,senta, sent,sz,sen,ent2,br1, n1,n2,n3}. Since the deceleration-acceleration cosmic transit is supported by recent observations \cite{11}, a specific form of the scale factor $a(t)$ that leads to a sign flipping in the evolution of the deceleration parameter $q$ from positive to negative can be utilized. Such hyperbolic scale factor ansatz has appeared in many cosmological contexts such as Bianchi cosmology \cite{pr}, and the cosmological models in the framework of Chern-Simons modified gravity \cite{sent,sent11}. In \cite{sen}, A. Sen introduced a Quintessence model using the same hyperbolic ansatz where the consistency of this ansatz with observations was the main motivation behind using it. It has also been shown that this hyperbolic form can also provide a unified description for cosmological evolution up to the late-time future \cite{sz}.
The deceleration and jerk parameters are respectively given as 
\begin{equation} \label{q1}
q(t)=-\frac{\ddot{a}a}{\dot{a}^2}=\frac{-\cosh^2(b t)+n}{\cosh^2(b t)},
\end{equation}
\begin{equation}\label{jerk}
j=\frac{\dddot{a}}{aH^3}= 1+\frac{2n^2-3n}{\cosh^2(b t)},
\end{equation}
The sign flipping of $q(t)$ is shown in Fig.1(a) for $n=\frac{1}{2}$ where $-1 \leq q(t)\leq 1$. The present value of $q(t)$ is supposed to be around $-0.55$ \cite{sz2}. De-Sitter expansion occurs at $q=-1$, power-law expansion happens for $-1 < q < 0$, and a super-exponential expansion occurs for $q<-1$. Using (\ref{q1}), the cosmic transit is supposed to occur at $q=0$ ( or $\ddot{a}=0$). Here we have,
\begin{equation}
t_{q=0}=\frac{1}{2 b}\ln(3+2\sqrt{2}),
\end{equation}
which gives $t\approx 0.88$ for $b=1$. The parameter $j(t)$ provides a convenient method to describe models close to $\Lambda$CDM \cite{jerk1,jerk2}. This parameter has the asymptotic value $j=1$ at late-time for the current flat model (figure \ref{e1}). For the most important papers that have shed the light on $q$ and $j$ constraints and on further terms beyond them see \cite{capo1, capo2, capo3, capo4, capo5, capo6, capo7, capo8, capo9, capo10}.
In this case, we get    

\begin{eqnarray}   
p(t)&=&\frac{1}{6144 \sinh^{6}(b t)\pi} \left((-384 b^2+512 \Lambda)\coth(bt)^6+(-72b^4+1152b^2-1536\Lambda)\cosh(bt)^4
\right. \\ \nonumber &+& \left.(-27b^6-72b^4-1152b^2+1536\Lambda)\coth(bt)^2+144b^4+384b^2-512\Lambda\right). \\
\rho(t)&=&\frac{1}{2048 \sinh(b t)^{6}\pi} \left((192 b^2-256 \Lambda)\coth(bt)^6+(-144b^4-384b^2+768\Lambda)\cosh(bt)^4
\right. \\ \nonumber &+& \left.(27b^6+144b^4+192b^2-768\Lambda)\coth(bt)^2+256\Lambda \right).\\
\omega(t)&=& \frac{1}{3}\left(\left(-384\left(b^2-\frac{4}{3} \Lambda\right)\cosh(bt)^6+(-72b^4+1152b^2-1536\Lambda)\cosh(bt)^4
\right. \right.\\ \nonumber &+& \left.\left.
(-27b^6-72b^4-1152 b^2+1536\Lambda )\cosh(bt)^2 +384b^{2}-512\Lambda +144b^{2}
\right)
\right)\\ \nonumber
&\div &\left(192\left(b^2-\frac{4}{3} \Lambda \right)\cosh(bt)^6+(-144b^4-384b^2+768\Lambda)\cosh(bt)^4
\right. \\ \nonumber &+& \left.
(-27b^6+144b^4+192 b^2-768\Lambda )\cosh(bt)^2+256\Lambda
\right).\\
\omega(z)&=&\left(-27A^{8}(1+(1+z)^{4}A^{4})(1+z)^{8}b^{6}-216(1+z)^{4}A^{4}m^{4}\left(\frac{1}{3}+(1+z)^{4}A^{4}
\right)
\right.\\ \nonumber &\times & \left.\
b^{4}-384b^{2}m^{4}+512\Lambda m^{4}
\right)
 \div \left(81 A^{8}(1+(1+z)^{4}A^{4})(1+z)^{8}b^{6}-432A^{4}
 \right.\\ \nonumber &\times & \left.\
 (1+(1+z)^{4}A^{4})(1+z)^{4}m^{2}b^{4}+(576+576(1+z)^{4}A^{4})m^{4}b^{2}-768\Lambda m^{4}
\right)
\end{eqnarray}

The last equation expresses the EoS parameter $\omega$ in terms of the redshift $z$. By calculating the limit of this equation as $z$ tends to $0$ (at the current epoch) we find that $\omega(z)=-1$ as predicted by observations, i.e. $\lim_{z\rightarrow 0} \omega(z)=-1 $ . The other quantities are plotted in Figure (\ref{cassimir55}). We have a physically acceptable behavior of the energy density $\rho(t)$ (\ref{F2}) and a future quintessence-dominated flat universe (\ref{u}). But the cosmic pressure $p(t)$ doesn't change the sign from positive to negative in corresponding to the cosmic transit. According to the dark energy assumption in which the negative pressure acts as a repulsive gravity, $p(t)$ should be positive in the early-time decelerating epoch and negative in the late-time accelerating epoch.

\subsection{Logamediate Inflation.}

Another interesting ansatz that has been proved to be consistent with CMB observations is the so called logamediate inflation scenario \cite{logscale} where the scale factor expands as $a(t)=e^{B\ln(t)^{\alpha}}$. Its consistency with observations is the main motivation behind using it in the current work. In this case, we get the jerk and deceleration parameters as

\begin{eqnarray}
j(t)&=&\frac{1}{B^2\alpha^2}\left[(-3\alpha+3)\ln(t)^{-2\alpha+1}+2\ln(t)^{-2\alpha+2}-3B\alpha \ln(t)^{-\alpha+1} \right. \\ \nonumber
&+& \left.(\alpha^2+3\alpha+2)\ln(t)^{-2\alpha}+(3B\alpha^2-3B\alpha)\ln(t)^{-\alpha}+B^2\alpha^2\right].\\
q(t)&=&{\frac { \left( -\alpha+1 \right) \ln(t)^{-\alpha}-B\alpha+ \ln(t)^{-\alpha+1}}{B\alpha}}~~~~~~~~~~~~~~~~~~~~~~~~~~~~~~~~~~
\end{eqnarray}

\begin{figure}[H]
  \centering            
  \subfigure[$q_1$]{\label{e}\includegraphics[width=0.32\textwidth]{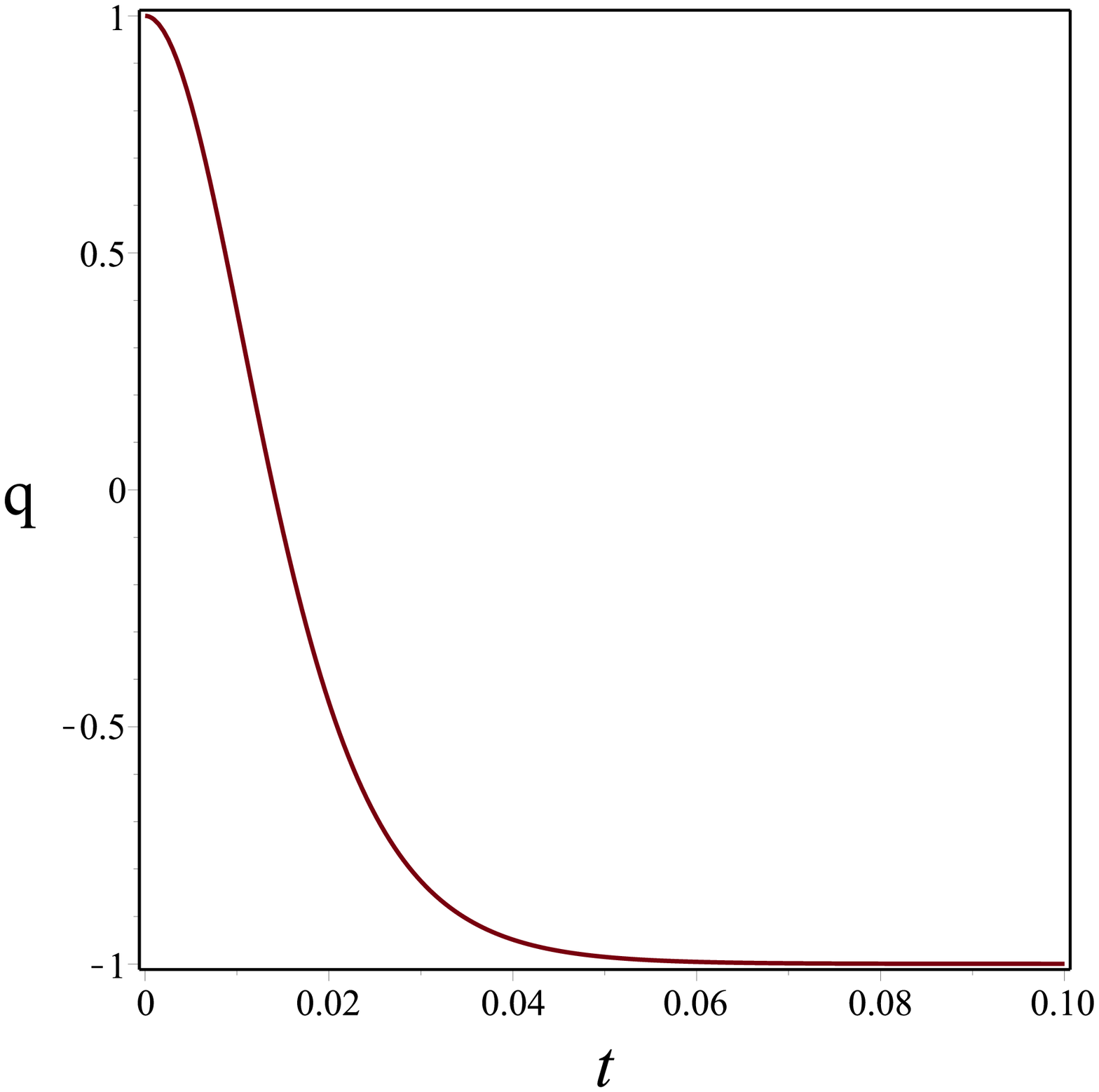}} 
\subfigure[$q_2$]{\label{ee}\includegraphics[width=0.32\textwidth]{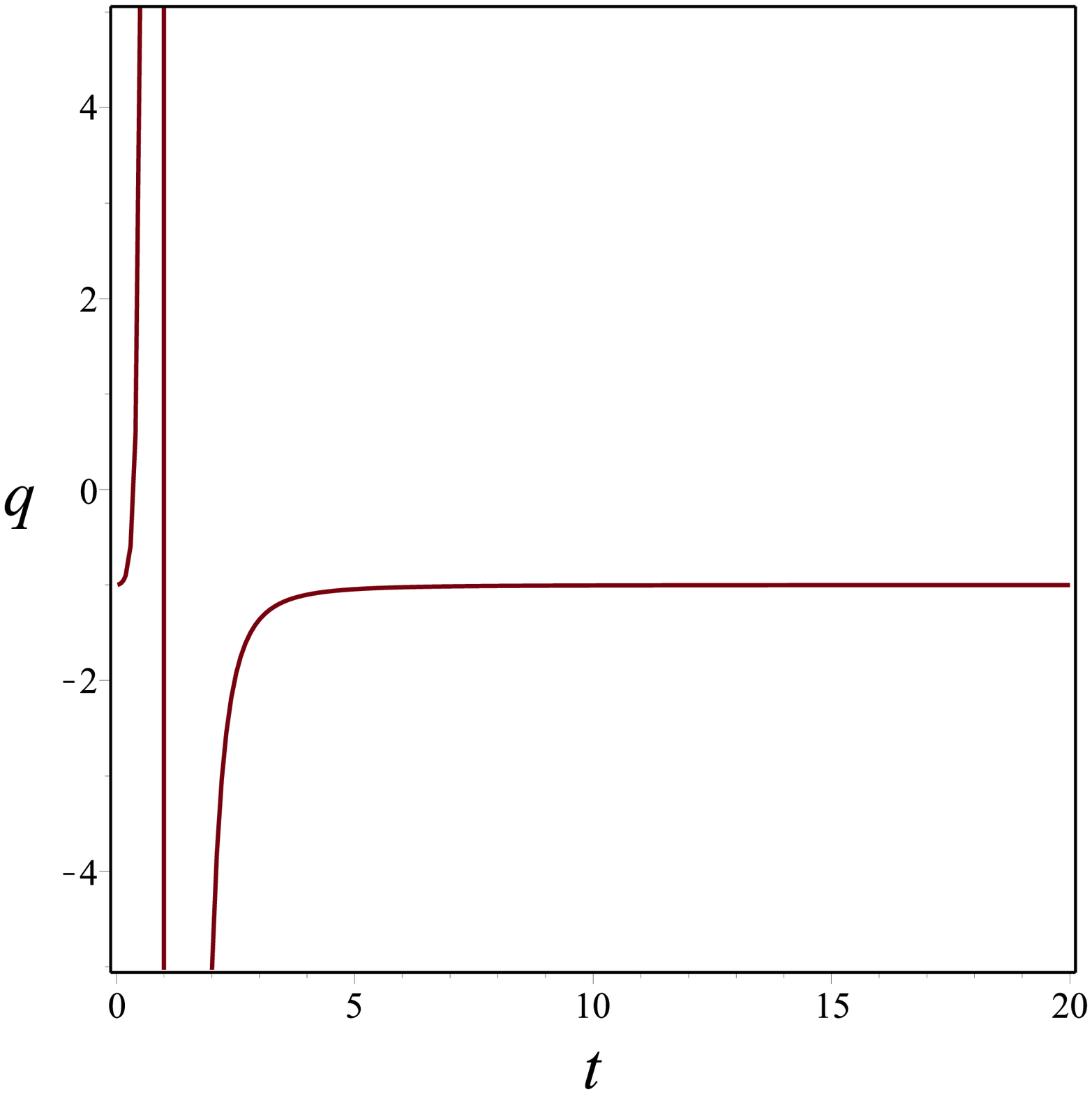}}
\subfigure[$q_3$]{\label{eee}\includegraphics[width=0.32\textwidth]{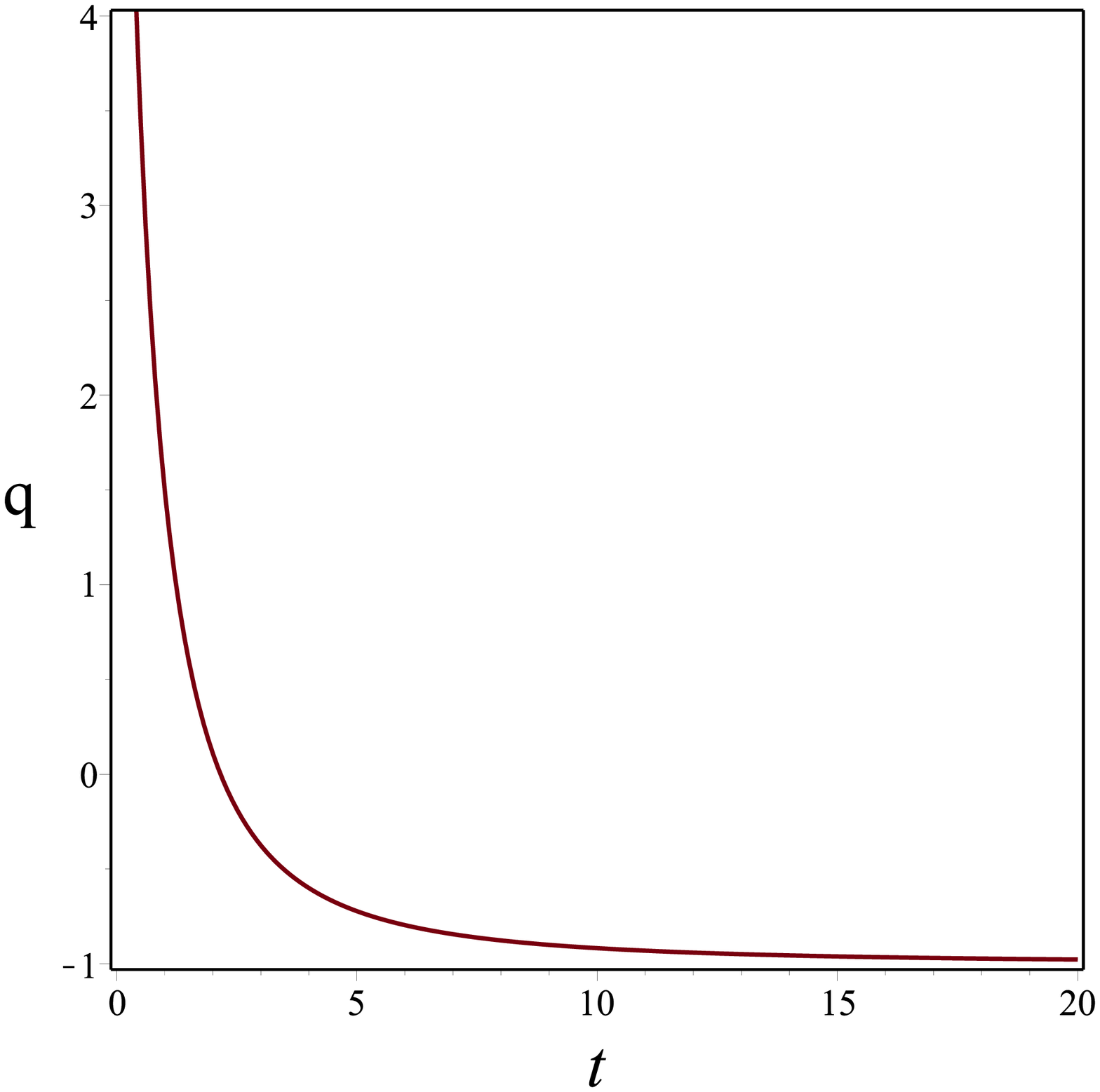}} \\
  \subfigure[$j_1$]{\label{e1}\includegraphics[width=0.32\textwidth]{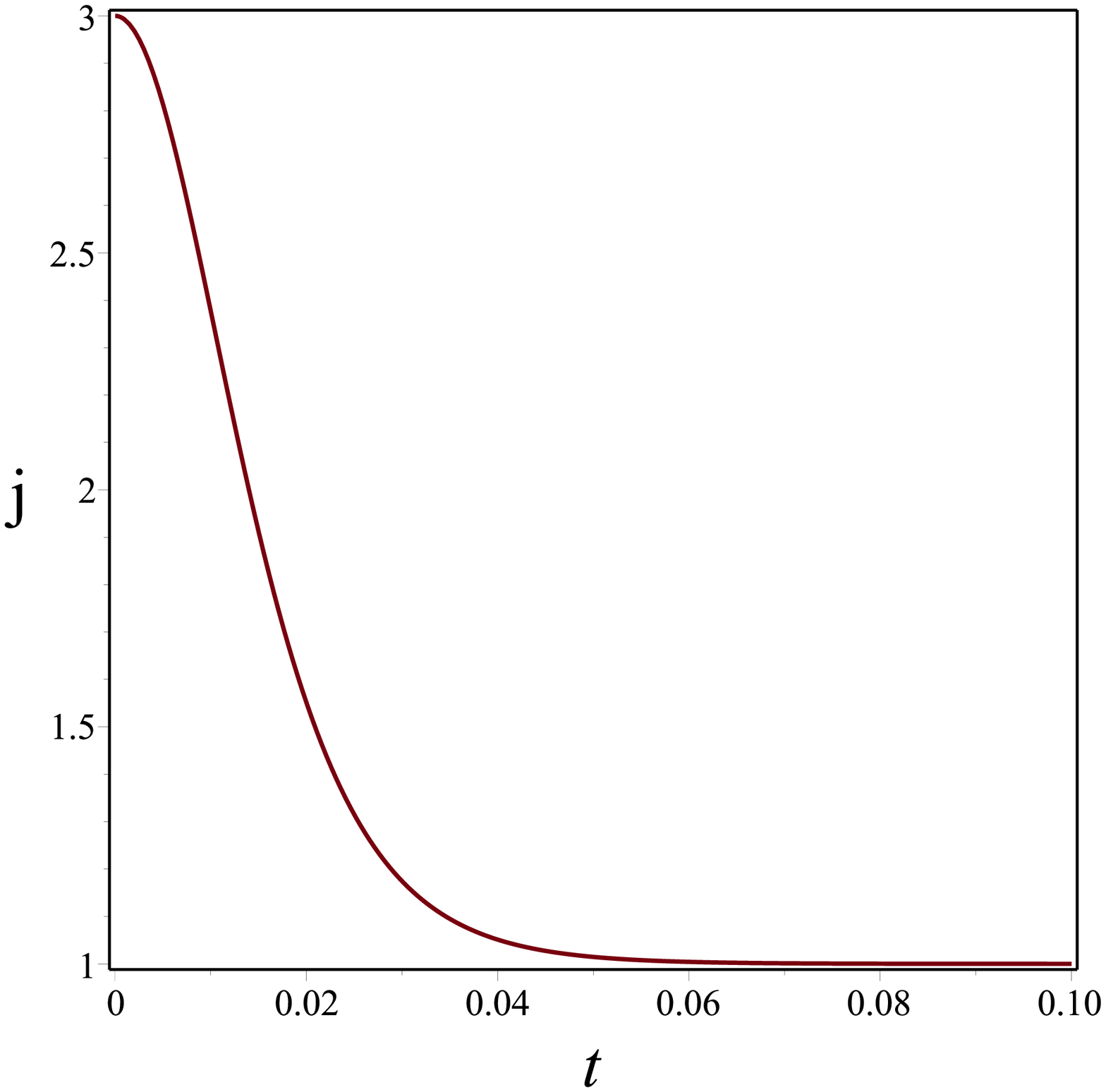}} 
\subfigure[$j_2$]{\label{ee1}\includegraphics[width=0.32\textwidth]{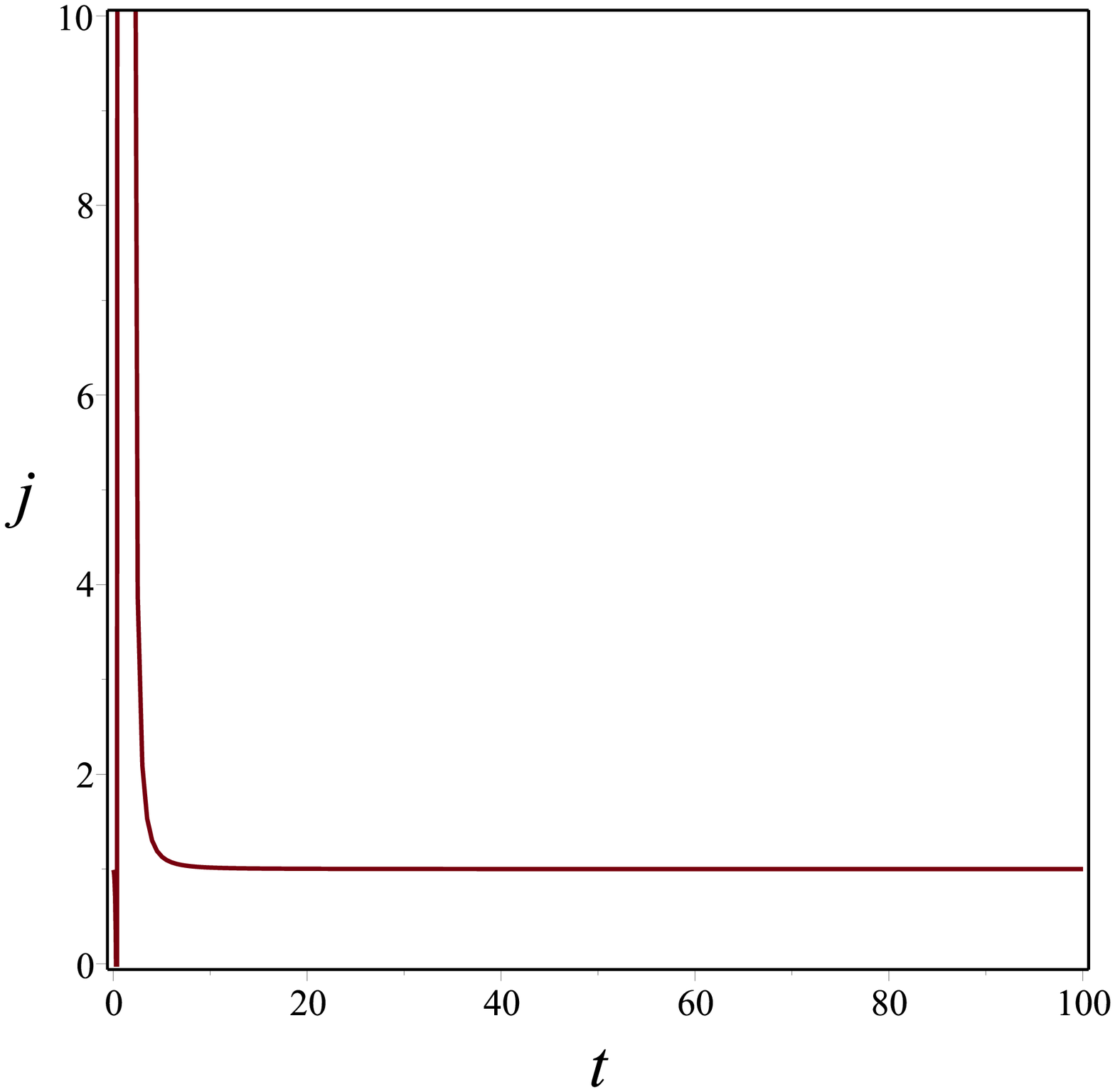}}
\subfigure[$j_3$]{\label{eee1}\includegraphics[width=0.32\textwidth]{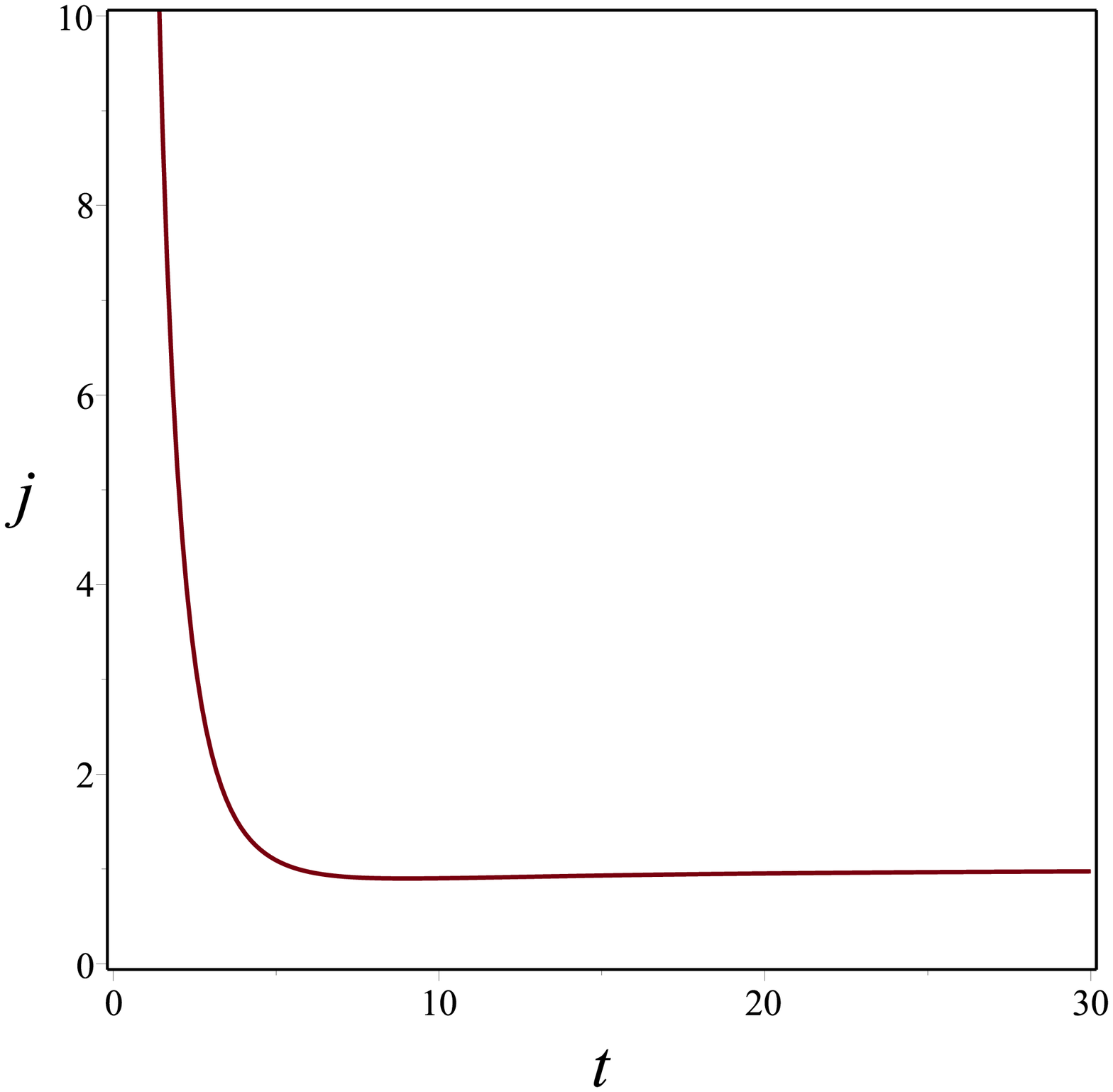}}
\caption{ The deceleration and jerk parameters for the hyperbolic, logamediate inflation and hybrid solutions labelled as $q_1$, $q_2$, $q_3$, $j_1$, $j_2$ and $j_3$ respectively. We have used these three ansatze mainly because of their consistency with observations. The cosmic deceleration-acceleration transit is allowed in all of them where there is a sign flipping of $q(t)$ from positive to negative. In addition, the behavior of the jerk parameter $j(t)$ at the current epoch represents another support for using those three ansatze. Since flat $\Lambda$CDM models have $j = 1$, the plots show that the jerk parameter tends to the flat $\Lambda CDM$ model for late-time in the three models. Here we have adopted the following numerical values: $n=\frac{1}{2}$, $A=0.1$, $b=63$, $\alpha=5$, $B=1$, and $\alpha_1=\beta=0.1$. }
  \label{55}
\end{figure}

Figures (\ref{ee}) and (\ref{ee1}) show the behavior of $q$ and $j$ of this solution. The deceleration parameter changes sign from negative (at the very early time) to positive, and then becomes negative again at late times which provides a more general description of cosmic expansion than the first solution. That means we have both an acceleration-deceleration cosmic transit which is expected to happen in the very early times, and a deceleration-acceleration cosmic transit which is supposed to happen at late times according to observations. (see \cite{vans} for a description of a similar scale factor). The pressure, energy density and Eos parameter can be expressed as 

\begin{eqnarray}\label{p}  
p(t)= \frac{1}{384}\frac{1}{\pi m^{4}t^{6}}\left(324\alpha^{2}(\alpha-1)\left(\left(^{2}m^{2}-\frac{13}{12}\right)\alpha-\frac{5}{3}t^{2}m^{2}+\frac{17}{12}\right)B^{2}\ln(t)^{-4+2\alpha}
\right. \\ \nonumber \left.
-27A^{2}\alpha^{2}(\alpha-1)^{2}(\alpha-2)^{2}\ln(t)^{(-6+2\alpha)}-162B^{3\alpha^{3}}(\alpha-2)(\alpha-1^{2})\ln(t)^{-6+3\alpha}
 \right.~ \\ \nonumber \left.
 +162B^{2}\alpha^{2}(\alpha-2)(\alpha-1)^{2}\ln(t)^{-5+2\alpha}+(324t^{2}m^{2}-810)\alpha^{3}(\alpha-1)B^{3}
\times\right. ~~~~~~~\\ \nonumber \left. 
 \ln(t)^{-4+3\alpha}
 -(862t^{2}m^{2}-324)\alpha^{2}(\alpha-1)B^{2}\ln(t)^{2\alpha-3}-243B^{4}\alpha^{4}(\alpha-1)^{2}
 \times\right. ~~~\\ \nonumber \left. 
 \ln(t)^{-6+4\alpha}
 +648\alpha^{3}(\alpha-1)B^{3}\left(-\frac{5}{4}+\alpha\right)\ln(t)^{-5+3\alpha}
  \right. ~~~~~~~~~~~~~~~~~~~~~~~~~~~~~~~ \\ \nonumber \left.
 -96\alpha^{2}\left(m^{4}t^{4}-\frac{45}{8}t^{2}m^{2}+\frac{9}{8}\right)B^{2}\ln(t)^{2\alpha-2}
 +486B^{4}\alpha^{4}(\alpha-1)\ln(t)^{-5+4\alpha}
  \right. ~~~~\\ \nonumber  \left.
+(-324t^{2}m^{2}+324)B^{3}\alpha^{3}\ln(t)^{3\alpha-3}
 -243A^{4}\ln(t)^{-4+4\alpha}\alpha^{4}
\right. ~~~~~~~~~~~~~~~~~~~~~~~~\\ \nonumber \left.
+36t^{2}\left(B\alpha(\alpha-1)(\alpha-2)(\alpha-3)\ln(t)^{-4+\alpha}-\frac{4}{3}B\alpha(\alpha-1)\left(t^{2}m^{2}-\frac{33}{4}\right)
\right.\right. ~~~\\ \nonumber \left.\left.
\ln(t)^{\alpha-2}
-6B\alpha(\alpha-1)(\alpha-2)\ln(n)^{\alpha-3}+\left(\frac{4}{3}t^{2}m^{2}-6\right)\alpha B\ln(t)^{\alpha -1}
\right.\right. ~~~~~~~~~\\ \nonumber \left.\left.
+\frac{8}{9}m^{2}t^{4}\lambda\right)
m^{2}\right)~~~~~~~~~~~~~~~~~~~~~~~~~~~~~~~~~~~~~~~~~~~~~~~~~~~~~~~~~~~~~~~~~~~~~~~~~~~.
\end{eqnarray}
\begin{eqnarray} \label{rho}
\rho(t)=\frac{1}{128}\frac{1}{\pi m^{4}t^{6}}\left(-72 \alpha^{2}(\alpha-1)B^{2}\left(\left(t^{2}m^{2}-\frac{39}{8}\right)\alpha-2t^{2}m^{2}+\frac{51}{8}\right)
\times  \right. \\ \nonumber \left.
\ln(t)^{-4+2\alpha}+27 B^{2}\alpha^{2}(\alpha-1)^{2}(\alpha-2)^{2}\ln(t)^{-6+2\alpha}+162B^{3}\alpha^{3}(\alpha-2)\times \right. \\ \nonumber \left.
(\alpha-1)^{2}\ln(t)^{-6+3\alpha}-162B^{2}\alpha^{2}(\alpha-2)(\alpha-1)^{2}\ln(t)^{-5+2\alpha}-216\alpha^{3}
\times  \right.  \\ \nonumber \left.
(\alpha-1)B^{3}\left(t^{2}m^{2}-\frac{15}{4}\right)\ln(t)^{-4+3\alpha}+216\alpha^{2}(\alpha-1)B^{2}
\times  \right. ~~~~~~~~~~~~~ \\ \nonumber \left.
\left(t^{2}m^{2}-\frac{3}{2}\right)\ln(t)^{2\alpha-3}+243B^{4}\alpha^{4}(\alpha-1)^{2}\ln(t)^{-6+4\alpha}-648\alpha^{3}
\times  \right. ~~~~ \\ \nonumber \left.
(\alpha-1)B^{3}\left(-\frac{5}{4}+\alpha\right)\ln(t)^{-5+3\alpha}+48 B^{2}\alpha^{2}\left(t^{2}m^{2}-\frac{3}{2}\right)^{2}
\times  \right. ~~~~~~~~ \\ \nonumber \left.
\ln(t)^{2\alpha-2}-486 B^{4}\alpha^{4}(\alpha-1)\ln(t)^{-5+4\alpha}+216 B^{3}\alpha^{3}\left(t^{2}m^{2}-\frac{3}{2}\right)
\times  \right. \\ \nonumber \left.
\ln(t)^{3\alpha-3}-16\lambda m^{4}t^{6}+243 B^{4}\ln(t)^{-4+4\alpha}\alpha^{4}\right).~~~~~~~~~~~~~~~~~~~~~~~~~~~~
\end{eqnarray}

\begin{eqnarray}
\omega(t)= \left(-162B^{2}\alpha^{2}(\alpha-2)(\alpha-1)^{2}\ln(t)^{1+2\alpha}-486\alpha^{4}B^{4}(\alpha-1)\ln(t)^{1+4\alpha}
  \right.  \\ \nonumber \left.
-324 B^{3}\alpha^{3}\left(t^{2}-\frac{5}{2}\right)(\alpha-1)\ln(t)^{2+3\alpha}+243\ln(t)^{2+4\alpha}\alpha^{4}B^{4}+324
\times  \right. ~ \\ \nonumber \left.
B^{3}\alpha^{3}(t-1)(t+1)\ln(t)^{3+3\alpha}-324 B^{2}\alpha^{2}(\alpha-1)\left(\left(t^{2}-\frac{13}{12}\right)\alpha
 \right.\right. ~~~~~~~ \\ \nonumber \left. \left.
-\frac{5}{3}t^{2}+\frac{17}{12}\right)\ln(t)^{2\alpha+2}+864B^{2}\left(t^{2}-\frac{3}{8}\right)\alpha^{2}(\alpha-1)\ln(t)^{2\alpha+3}
\right. ~~~~~~~~ \\ \nonumber \left.
+96 B^{2}\alpha^{2}\left(t^{4}-\frac{45}{8}t^{2}+\frac{9}{8}\right)\ln(t)^{2\alpha+4}+27B^{2}\alpha^{2}(\alpha-1)^{2}(\alpha-2^{2})
\times \right.  ~~ \\ \nonumber \left.
\ln(t)^{2\alpha}+162B^{3}\alpha^{3}(\alpha-2)(\alpha-1)^{2}\ln(t)^{3\alpha}+243\alpha^{4}B^{4}(\alpha-1)^{2}
\times \right. ~~ ~~~ \\ \nonumber \left.
\ln(t)^{4\alpha}+216B\alpha t^{2}(\alpha-1)(\alpha-2)\ln(t)^{3+\alpha}-36B\alpha t^{2}(\alpha-1)
\times \right.   ~~ ~~~~~ \\ \nonumber \left.
(\alpha-2)(\alpha-3)\ln(t)^{\alpha+2}+48B\alpha(\alpha-1)\left(t^{2}-\frac{33}{4}\right)t^{2}\ln(t)^{\alpha+4}
 \right.   ~~~ ~~~~ \\ \nonumber \left.
-48B\alpha\left(t^{2}-\frac{9}{2}\right)t^{2}\ln(t)^{\alpha+5}-32\Lambda t^{6}ln(t)^{6}-648B^{3}\alpha^{3}(\alpha-1)
\times \right.   ~~~~ \\ \nonumber \left.
\left(-\frac{5}{4}+\alpha\right)\ln(t)^{3\alpha+1} 
\right)\div\left(216B^{2}\left(\left(t^{2}-\frac{39}{8}\right)\alpha -2t^{2}+\frac{51}{8}\right)\alpha^{2}
\times \right.    \\ \nonumber \left.
(\alpha-1)\ln (t)^{2\alpha+2}+486B^{2}\alpha^{2}(\alpha-2)(\alpha-1)^{2}\ln(t)^{1+2\alpha}+648B^{3}\alpha^{3}
\times \right.    \\ \nonumber \left.
\left(t^{2}-\frac{15}{4}\right)(\alpha-1)\ln(t)^{2+3\alpha}-648B^{2}\alpha^{2}(\alpha -1)\left(t^{2}-\frac{3}{2}\right)\ln(t)^{2\alpha+3}
\times \right.    \\ \nonumber \left.
-144A^{2}\alpha^{2}\left(t^{2}-\frac{3}{2}\right)^{2}\ln(t)^{2\alpha+4}+1944 B^{3}\alpha^{3}(\alpha-1)\left(-\frac{5}{4}+\alpha\right)
\times \right.  ~~  \\ \nonumber \left.
\ln(t)^{3\alpha+1}+1458 \alpha^{4}B^{4}(\alpha-1)\ln(t)^{1+4\alpha}-648 B^{3}\alpha^{3}\left(t^{2}-\frac{3}{2}\right)\ln(t)^{3+3\alpha}
 \right.    \\ \nonumber \left.
-729\ln(t)^{2+4\alpha}\alpha^{4}B^{4}-81 B^{2}\alpha^{2}(\alpha-1)^{2}(\alpha-2)^{2}\ln(t)^{2\alpha}-486 B^{3}
 \times \right.   ~~~~ \\ \nonumber \left.
\alpha^{3}(\alpha-2)(\alpha-1)^{2}\ln(t)^{3\alpha}
-729 \alpha^{4}B^{4}(\alpha-1)^{2}\ln(t)^{4\alpha}+48\Lambda t^{6}\ln(t)^{6}
\right)
\end{eqnarray}

Figure (\ref{cassimir55}) shows a physically acceptable behavior of the energy density $\rho(t)$. We can see that, as in the first solution, the cosmic pressure $p(t)$ doesn't change sign from positive to negative during its evolution in corresponding to the deceleration-acceleration cosmic transit. Because cosmic transit already exists in the model (Figur \ref{ee}), this behavior of $p(t)$ which is always negative doesn't help to explain this cosmic transit. In addition to this incomplete behavior of cosmic pressure, the evolution of the EoS parameter $\omega(t)$ also shows a singularity at early-time. For late-times, $\omega(t)$ has the asymptotic value $\approx -\frac{1}{3}$ which is exactly the same asymptotic value of the first solution. So, both solutions predict a future quintessence-dominated universe and also suffer from a divergence at some point.

\subsection{Hybrid solution.}

Another ansatz which leads to a deceleration-acceleration cosmic transit with the jerk parameter tends to the flat $\Lambda CDM$ model for late-time is the hybrid scale factor \cite{hyb1, kumar}. This consistency with observations is the main motivation behind using it. 
The hybrid ansatz is a mixture of both power-law and exponential-law cosmologies, the scale factor expands as \cite{hyb1, kumar}: 
\begin{equation} \label{hyb2}
a(t)=a_1 t^{\alpha_1} e^{\beta t},
\end{equation}
where $a_1>0$, $\alpha_1 \geq 0$ and $\beta \geq 0$ are constants.  For $\beta=0$ we get the power-law cosmology, and for $\beta=0$ we obtain the exponential-law cosmology. New cosmologies can be investigated for $\alpha_1>0$ and $\beta>0$. In this case we get the deceleration and jerk parameters as

\begin{eqnarray}
q(t)&=&-\frac{\ddot{a}a}{\dot{a}^2}=\frac{\alpha_1}{(\beta t+\alpha_1)^2}-1  \\  
j(t)&=&{\frac {{\alpha_1}^{3}+ \left( 3\,\beta\,t-3 \right) {\alpha_1}^{2}+
 \left( 3\,{\beta}^{2}{t}^{2}-3\,\beta\,t+2 \right) \alpha_1+{\beta}^{3}
{t}^{3}}{ \left( \beta\,t+\alpha_1 \right) ^{3}}}
\end{eqnarray}
The cosmic transit occurs at $t=\frac{\sqrt{\alpha_1}-\alpha_1}{\beta}$ which restricts $\alpha_1$ in the range $0<\alpha_1<1$ \cite{kumar}. The pressure, energy density, and EoS parameter are given as 
\begin{eqnarray}\label{p3} 
p(t)=\frac{1}{384} \frac{1}{m^{4}\pi t^{6}}\left(-243\alpha_1^{4}+(-324m^{2}t^{2}-486\beta t+324)\alpha_1^{3}+(-96m^{4}t^{4}
 \right.  \\ \nonumber - \left.
648m^{2}t^{3}\beta -108+(-243\beta^{2}+540m^{2})t^{2}+324\beta t)\alpha_1^{2}-192m^{2}
 \times \right.  ~~ \\ \nonumber \left.
  \left(\frac{9}{8}+m^{2}t^{3}\beta +\left(-\frac{1}{4}m^{2}+\frac{27}{16}\beta^{2}\right)t^{2}-\frac{9}{4}\beta t\right)t^{2}\alpha_1 -96m^{4}\left(\beta^{2}-\frac{1}{3}\Lambda\right)t^{6}
\right)
\end{eqnarray}

\begin{eqnarray} \label{rho3}
\rho(t)&=&\frac{1}{128 m^{4}\pi t^{6}}\left(243\alpha_1^{4}+(216m^{2}t^{2}+486\beta t-324)\alpha_1^{3}+(48m^{4}t^{4}+432m^{2}t^{3}\beta
 \right.    \\ \nonumber  &+& \left.
 108+(243\beta^{2}-144m^{2})t^{2}-324\beta t)\alpha_1^{2}+96m^{2}\left(m^{2}t^{2}+\frac{9}{4}\beta t -\frac{3}{2}\right)
 \right.      \\ \nonumber &\times&   \left.
  \beta t^{3}\alpha_1 +48m^{4}\left(\beta^{2}-\frac{1}{3}\Lambda\right)t^{6}
 \right) ~~~~~~~~~~~~~~~~~~~~~~~~~~~~~~~~~~~~
\end{eqnarray}

\begin{eqnarray}
\omega (t)=\left(-243\alpha_1^{4}+(-324m^{2}t^{2}-486\beta t+324)\alpha_1^{3}+(-96m^{4}t^{4}-648m^{2}t^{3}\beta
 \right.      \\ \nonumber  -  \left.
108+(-243\beta^{2}+540m^{2})t^{2}+324\beta t)\alpha_1^{2}-192m^{2}\left(\frac{9}{8}+m^{2}t^{3}\beta +
  \right.   \right.  ~~~~ \\ \nonumber +  \left. \left.
 \left(-\frac{1}{4}m^{2}+\frac{27}{16}\beta ^{2}\right)t^{2}-\frac{9}{4}\beta t\right)t^{2}\alpha_1 -96m^{4}\left(\beta^{2}-\frac{1}{3}\Lambda \right)t^{6}\right)\div \left(729
  \right.   ~~~   \\ \nonumber \times  \left.
  \alpha_1^{4}+(648m^{2}t^{2}+1458\beta t-972)\alpha_1^{3}+(324+144m^{4}t^{4}_1296m^{2}t^{3}\beta 
   \right.   ~~~~~~   \\ \nonumber + \left. 
  (729\beta^{2}-432m^{2})t^{2}-972\beta t)\alpha_1^{2}+288m^{2}\left(m^{2}t^{2}+\frac{9}{4}\beta t-\frac{3}{2}\right)\beta t^{3}\alpha_1
 \right.   ~~   \\ \nonumber + \left. 
144 m^{4}\left(\beta^{2}-\frac{1}{3}\Lambda\right) t^{6}
 \right)~~~~~~~~~~~~~~~~~~~~~~~~~~~~~~~~~~~~~~~~~~~~~~~~~~~~~~~~~
\end{eqnarray}

\begin{figure}[H]
  \centering            
  \subfigure[$p_1$]{\label{F63}\includegraphics[width=0.3\textwidth]{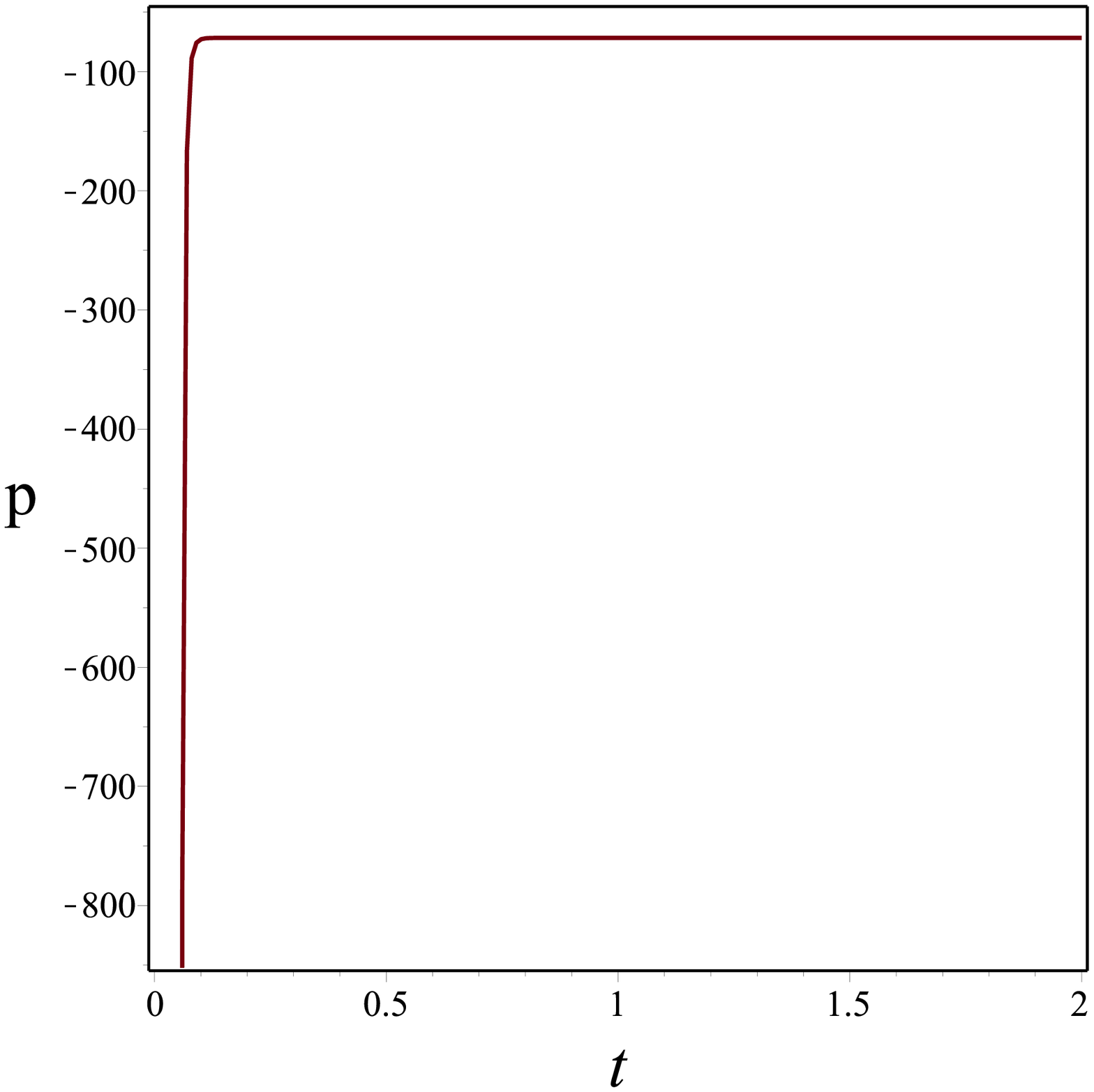}} 
\subfigure[$p_2$]{\label{F6202y28}\includegraphics[width=0.3\textwidth]{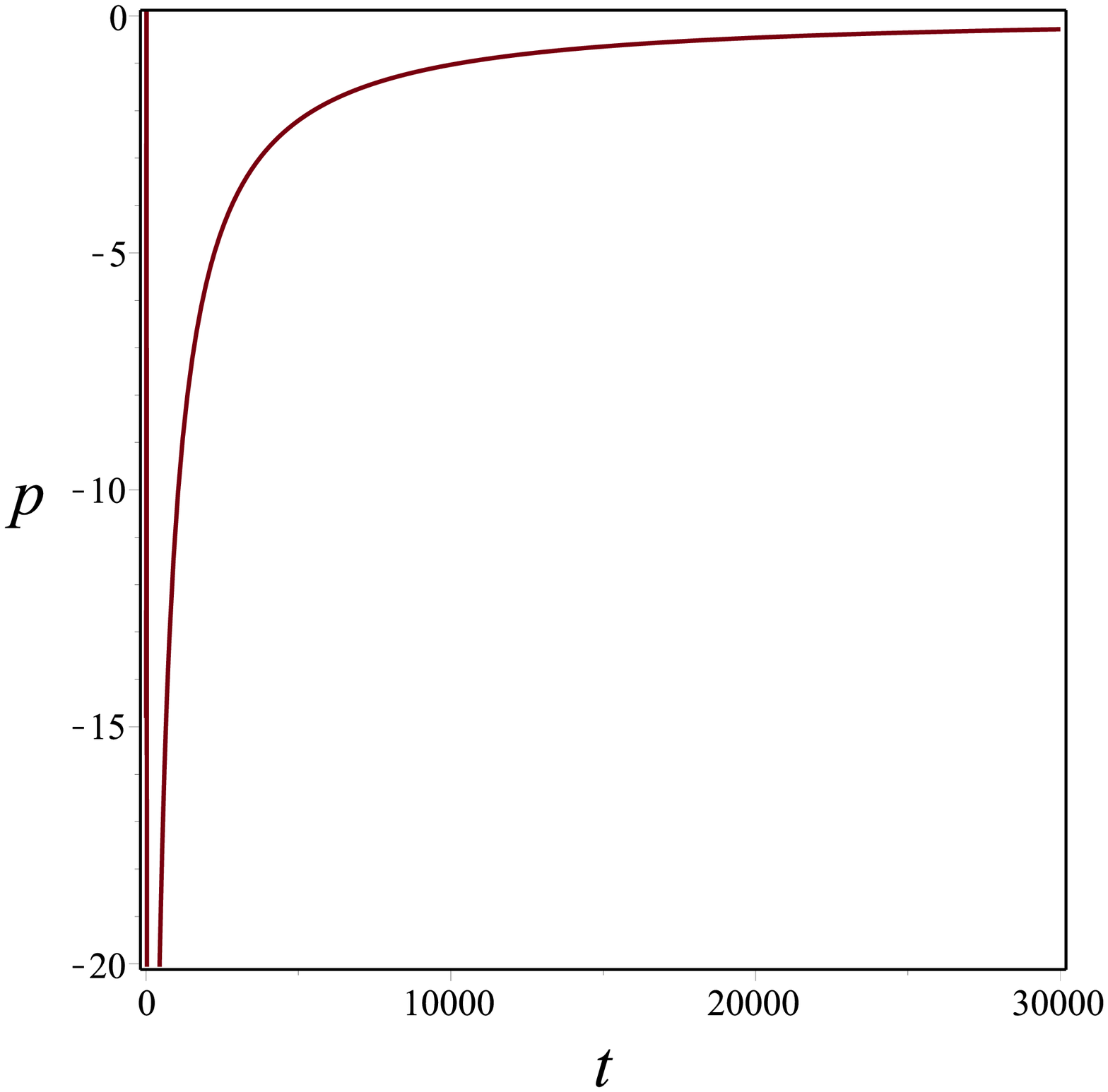}}
\subfigure[$p_3$]{\label{F622}\includegraphics[width=0.3\textwidth]{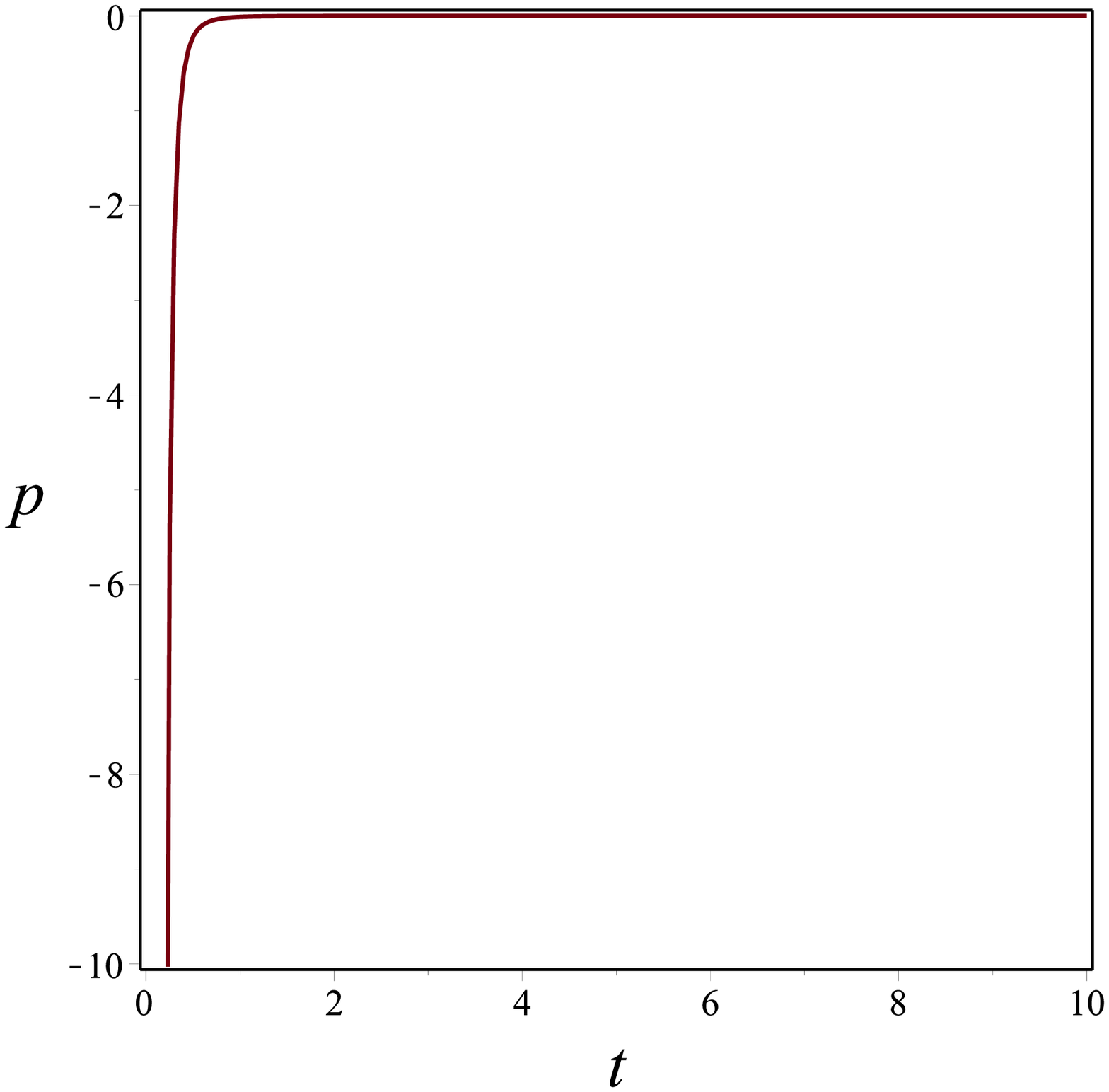}} \\
 \subfigure[$\rho_{1}$]{\label{F2}\includegraphics[width=0.3\textwidth]{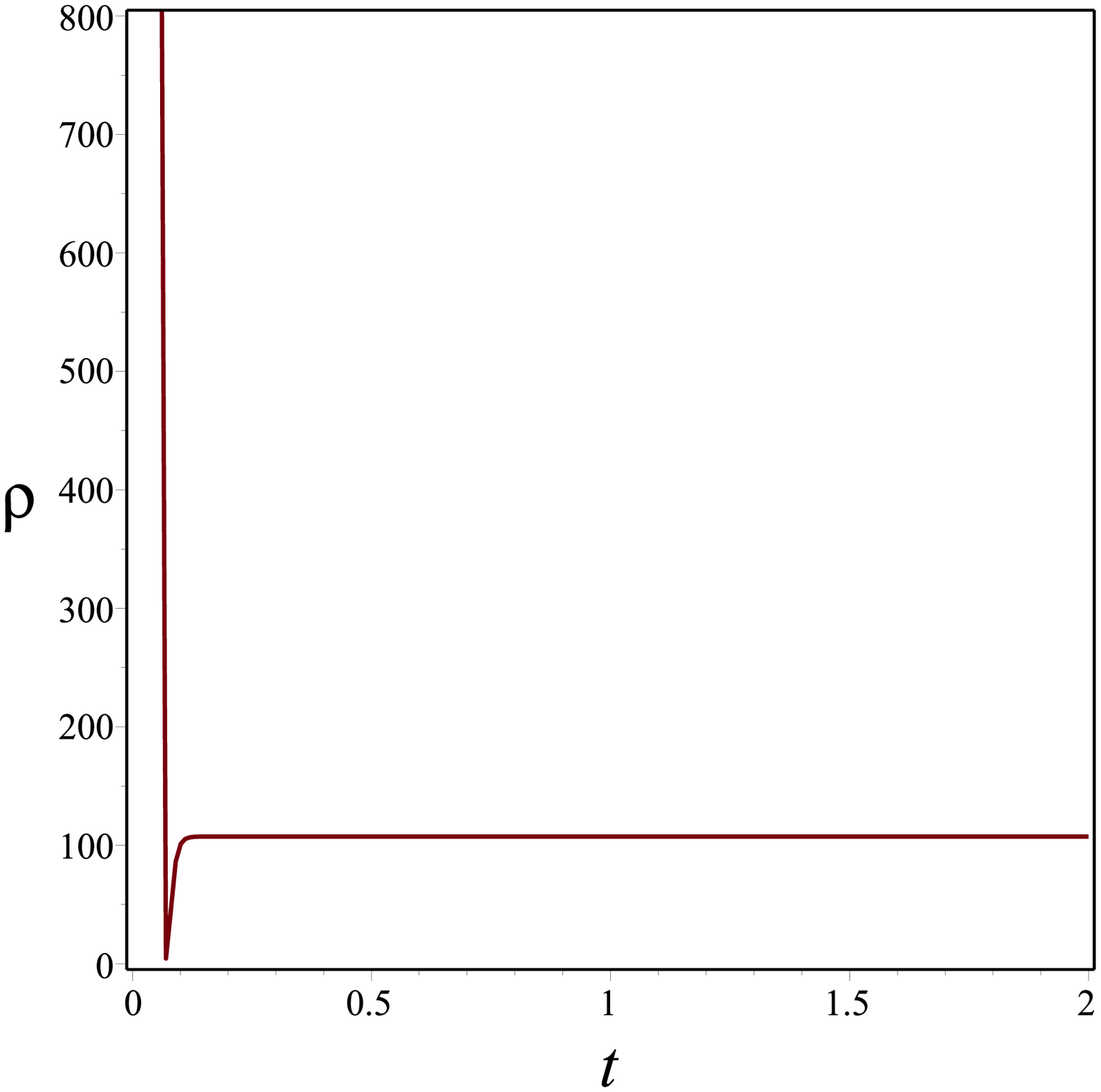}}
 \subfigure[$\rho_{2}$]{\label{F62262}\includegraphics[width=0.3\textwidth]{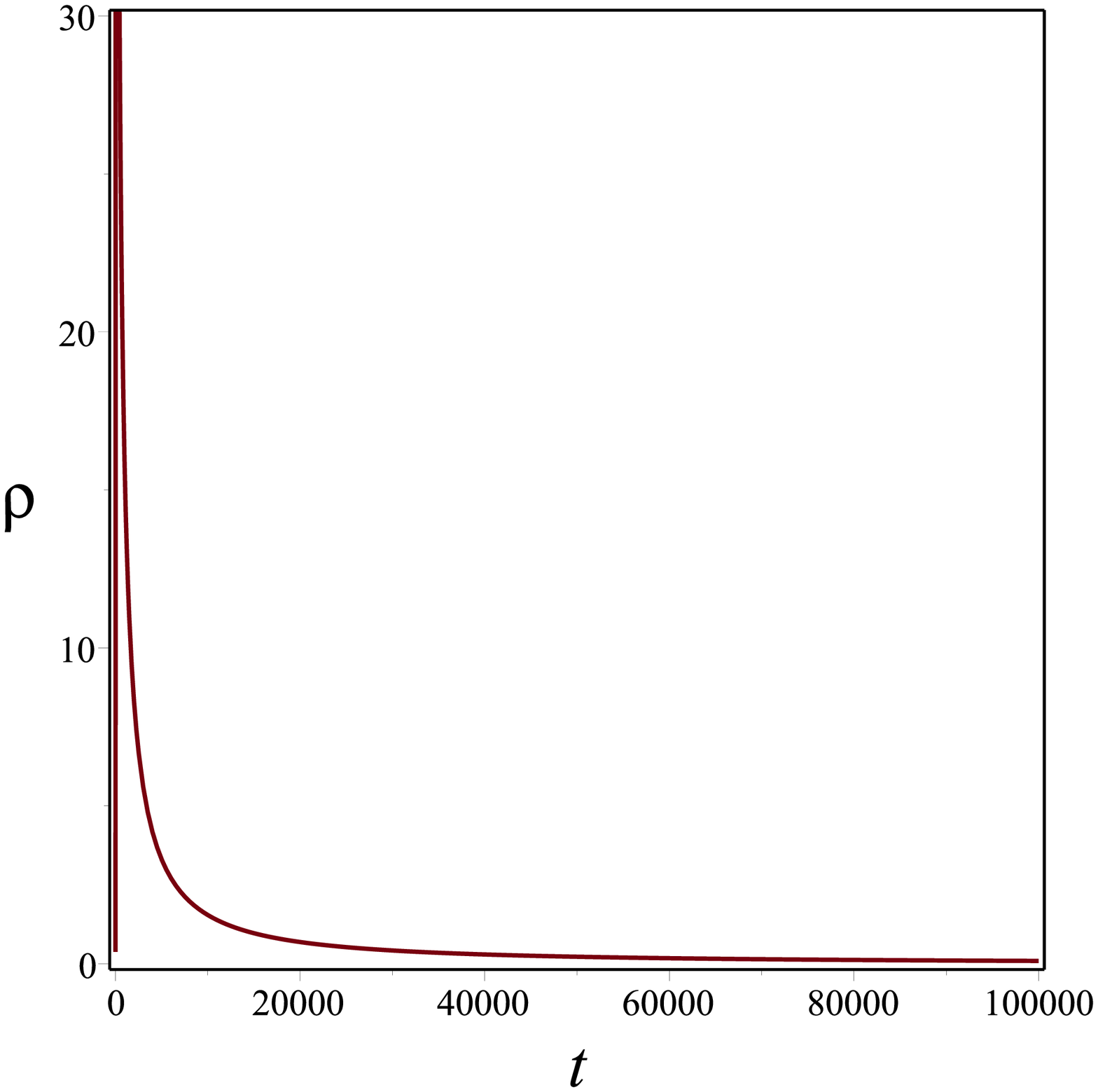}}
 \subfigure[$\rho_{3}$]{\label{F29}\includegraphics[width=0.3\textwidth]{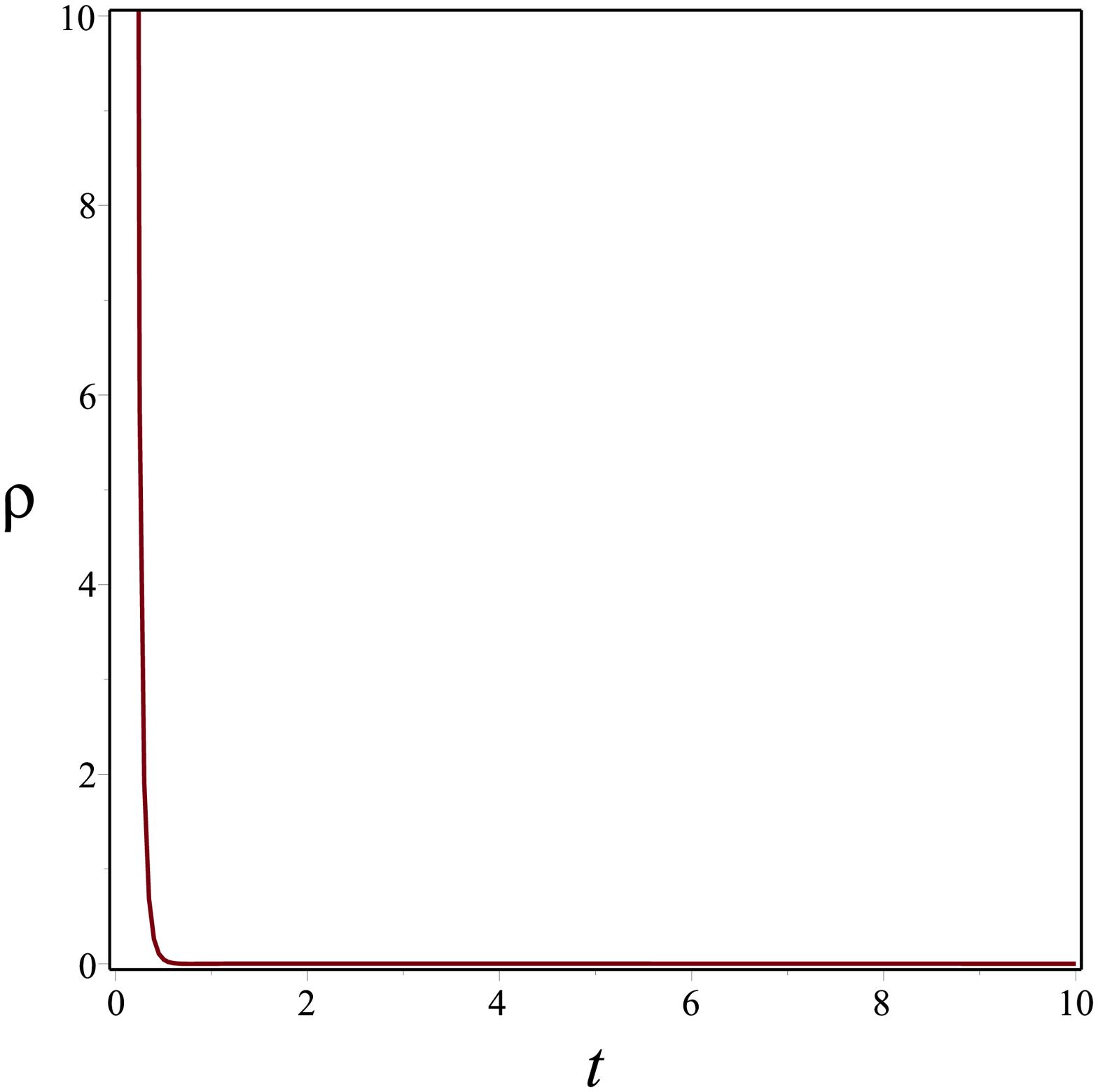}}\\
 \subfigure[$\omega_{1}$]{\label{u}\includegraphics[width=0.3\textwidth]{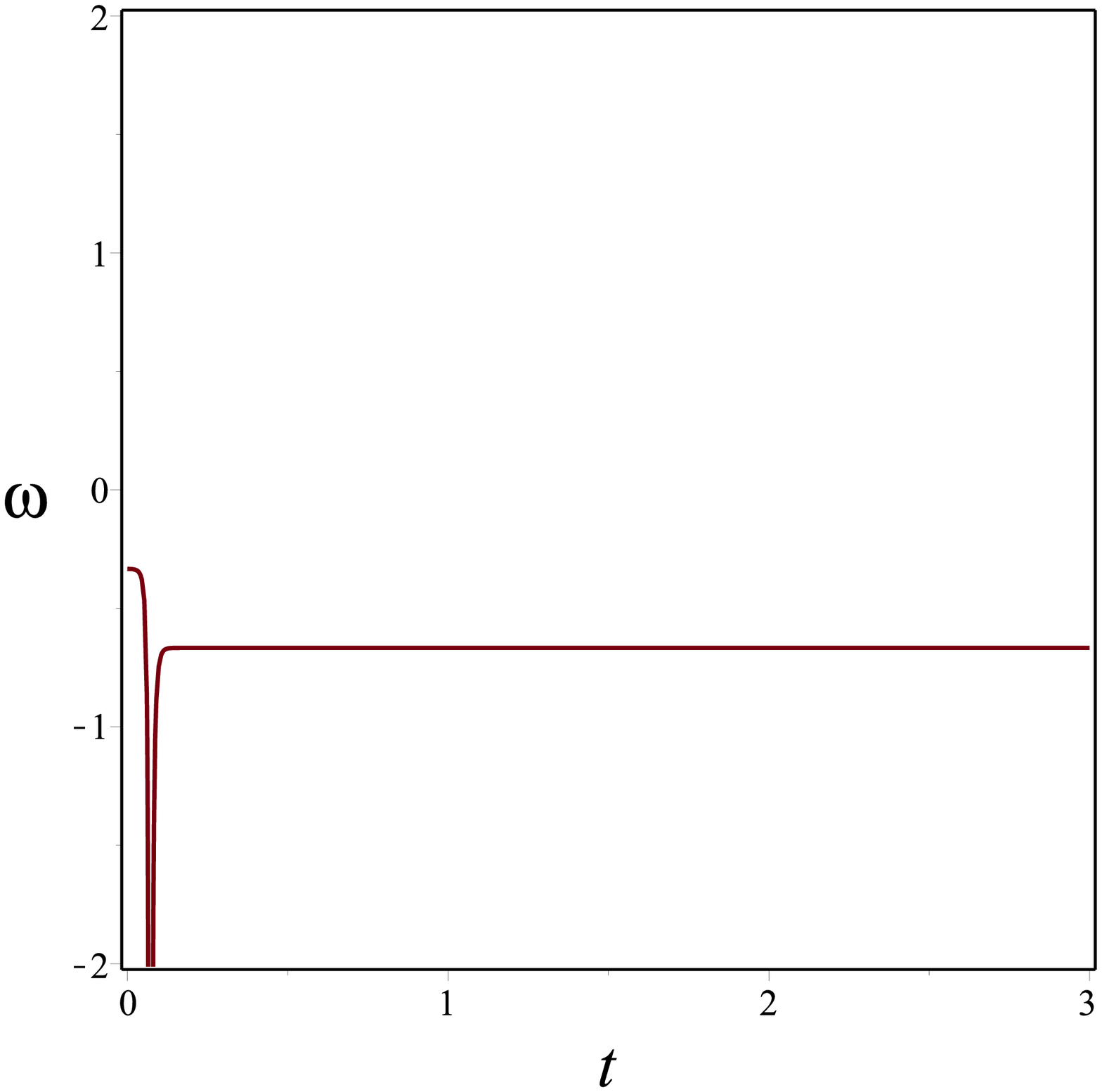}}
 \subfigure[$\omega_{2}$]{\label{j}\includegraphics[width=0.3\textwidth]{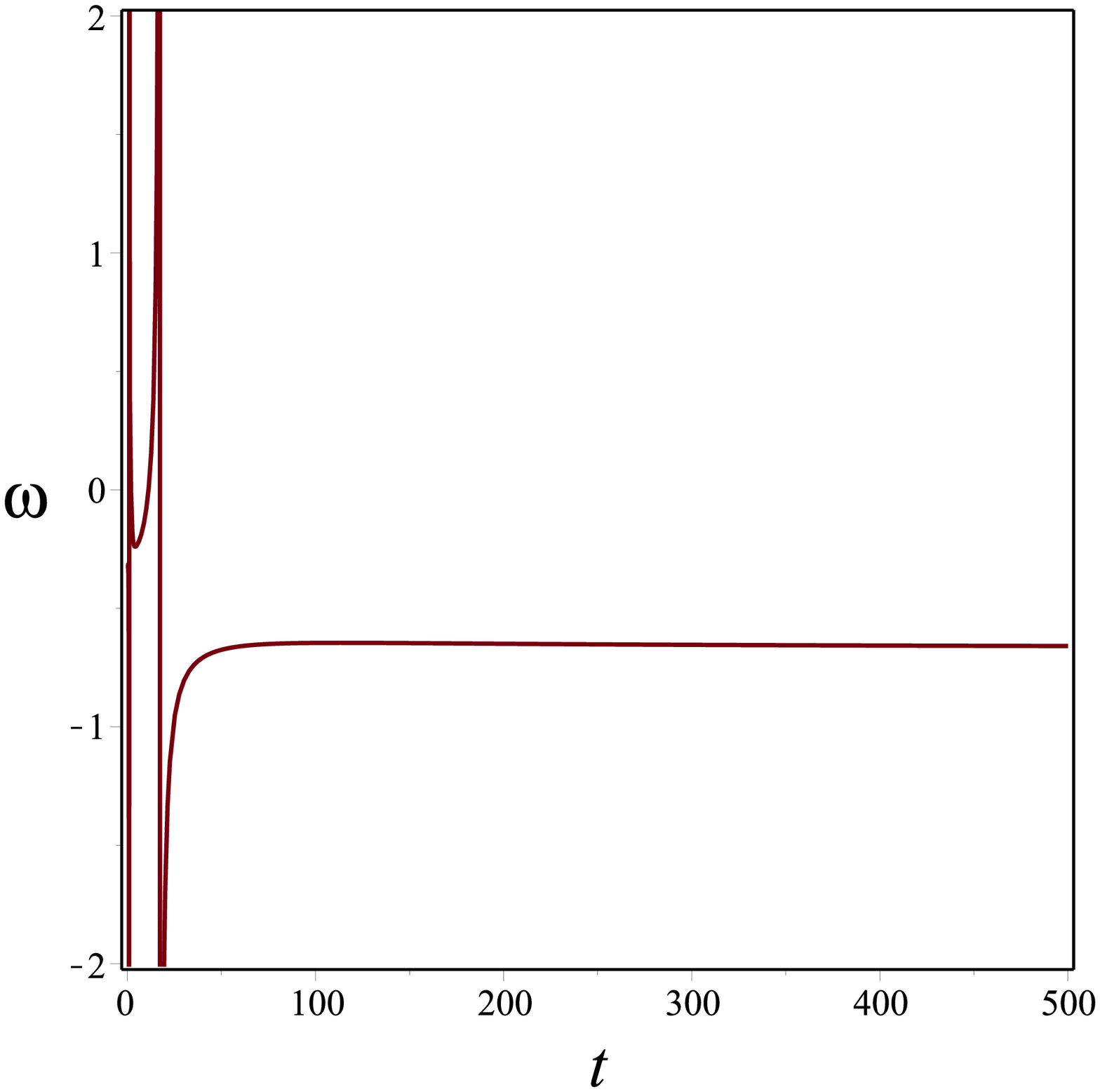}}
 \subfigure[$\omega_{3}$]{\label{F20}\includegraphics[width=0.3\textwidth]{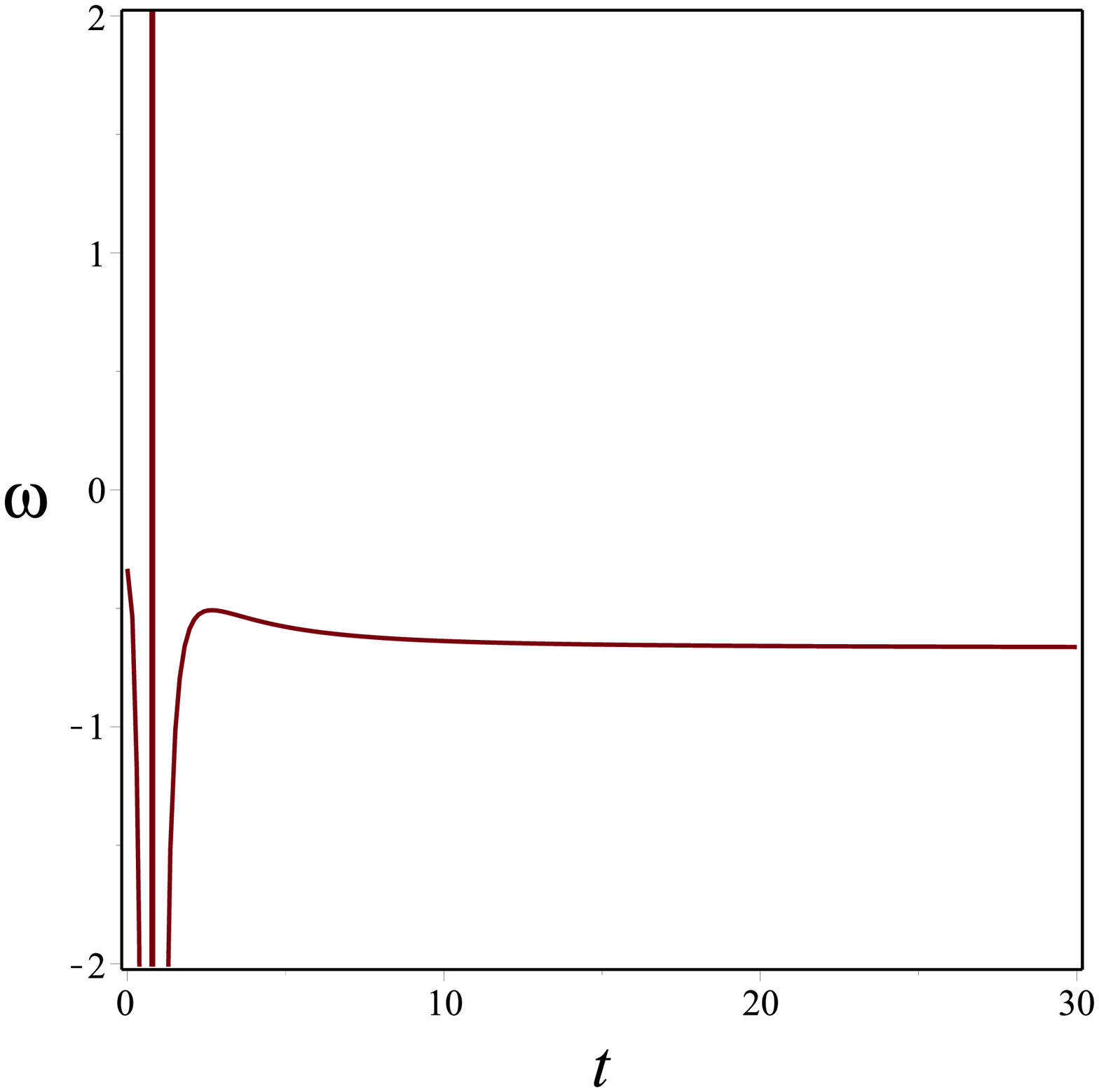}}
\caption{The pressure, energy density and EoS parameter for the three solutions. We have a physically accepted behavior of the energy density where $\rho \rightarrow + \infty$ as $t \rightarrow 0$. However, for the three different ansatze, the scale-invariance cosmology fails to provide a complete behavior of $p(t)$ and $\omega(t)$. Since dark energy has a negative pressure effect, there should be a positive-to-negative sign flipping in cosmic pressure in corresponding to the positive-to-negative sign flipping of the deceleration parameter (Figure \ref{55}). We also observe a singularity in the behavior of $\omega(t)$ which means that a complete description for the dark energy evolution can't be obtained. While the three forms of $a(t)$ lead to a complete description of $p(t)$ and $\omega(t)$ in other theories, the failure to predict a complete behavior of $p(t)$ and $\omega(t)$ using the same forms in the scale-invariance cosmology puts the cosmological viability of the theory under dispute. Here we have adopted the following numerical values: $n=\frac{1}{2}$, $A=0.1$, $b=63$, $\alpha=5$, $B=1$, $\alpha_1=\beta=0.1$, $m=1$, and $\Lambda=0.001$. We have tried different very small positive values of $\Lambda$ and found no change in the behavior of the plots.}
  \label{cassimir55}
\end{figure}

Here also we obtain the same behavior we have got in the previous two solutions for $p(t)$, $\rho(t)$ and $\omega(t)$. Figure (\ref{cassimir55}) shows a physically acceptable behavior of $\rho(t)$, a singularity in the evolution of the EoS parameter $\omega(t)$, and no sign flipping of $p(t)$ from positive to negative in corresponding to the deceleration-acceleration cosmic transit.

\section{Comparison with other theories}

In the previous section, we have studied three different toy models in the framework of a scale-invariance cosmology where $p(t)$ and $\omega(t)$ always show undesirable and incomplete behavior. This result makes the ability of the scale-invariance cosmology to provide a complete description for dark energy evolution uncertain. As we have mentioned in the introduction section, the three empirical forms of $a(t)$ used in the current study have been utilized in other modified gravity theories by many researchers where better behaviors of both $p$ and $\omega$ have been obtained. The positive-to-negative sign flipping of cosmic pressure helps to explain the cosmic transition from deceleration to acceleration. In this section, we give some examples of those works where the same forms of $a(t)$ have been used in different cosmological contexts. \par 

\begin{figure}[H]
  \centering            
  \subfigure[$p$]{\label{gg}\includegraphics[width=0.32\textwidth]{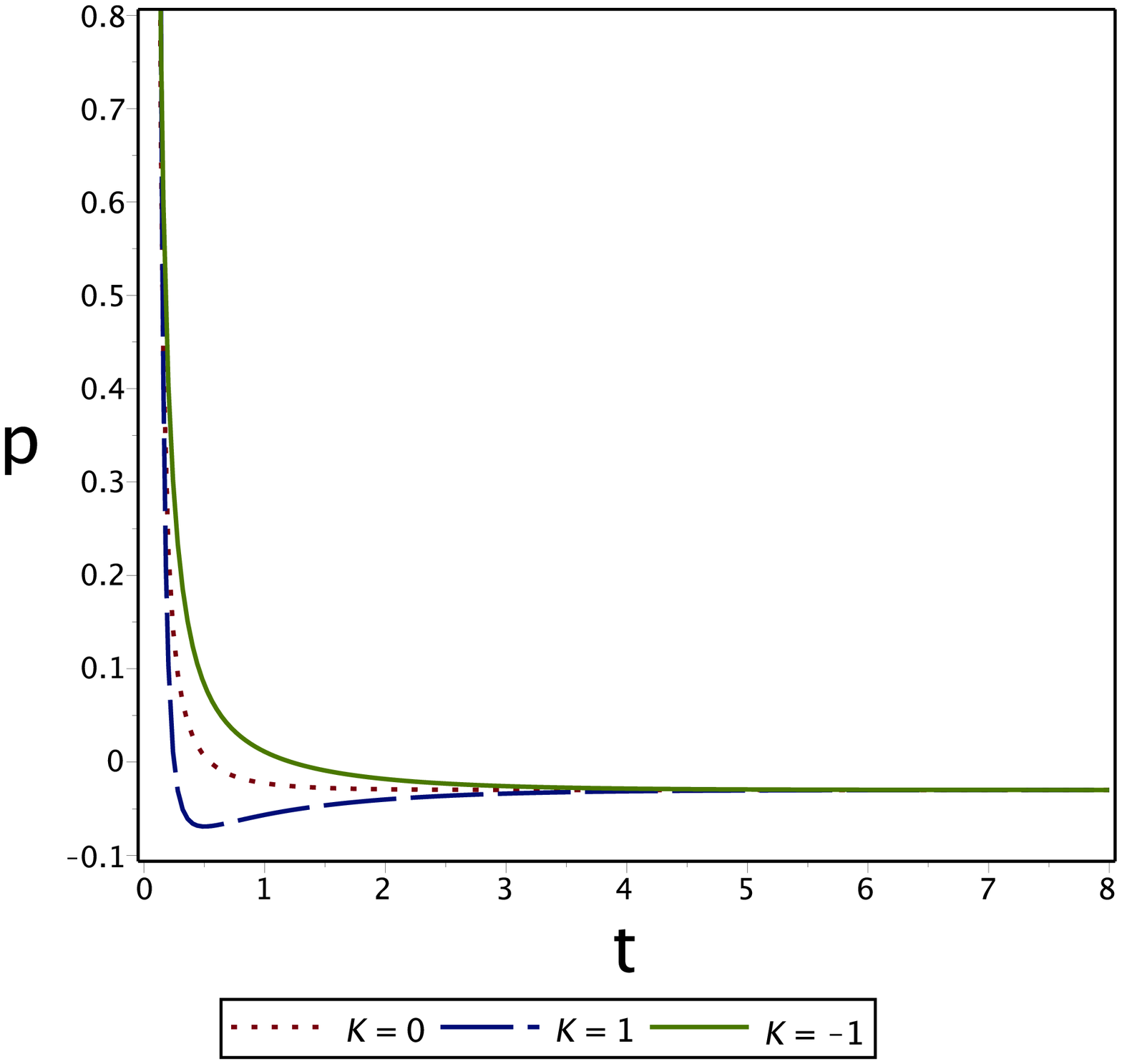}} 
	\subfigure[$\omega$]{\label{ggg}\includegraphics[width=0.32\textwidth]{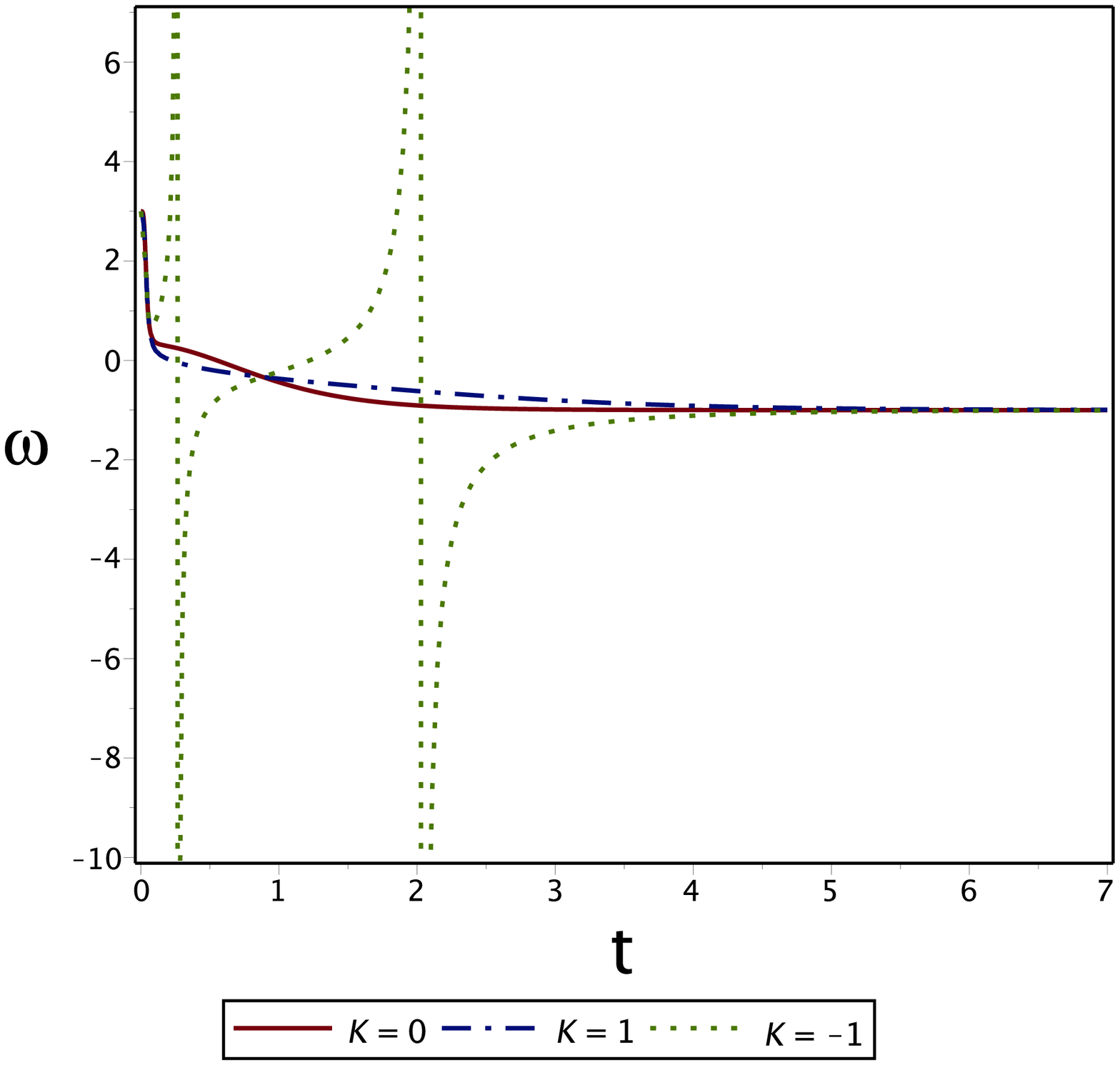}}
	 \subfigure[$p$]{\label{gg1}\includegraphics[width=0.32\textwidth]{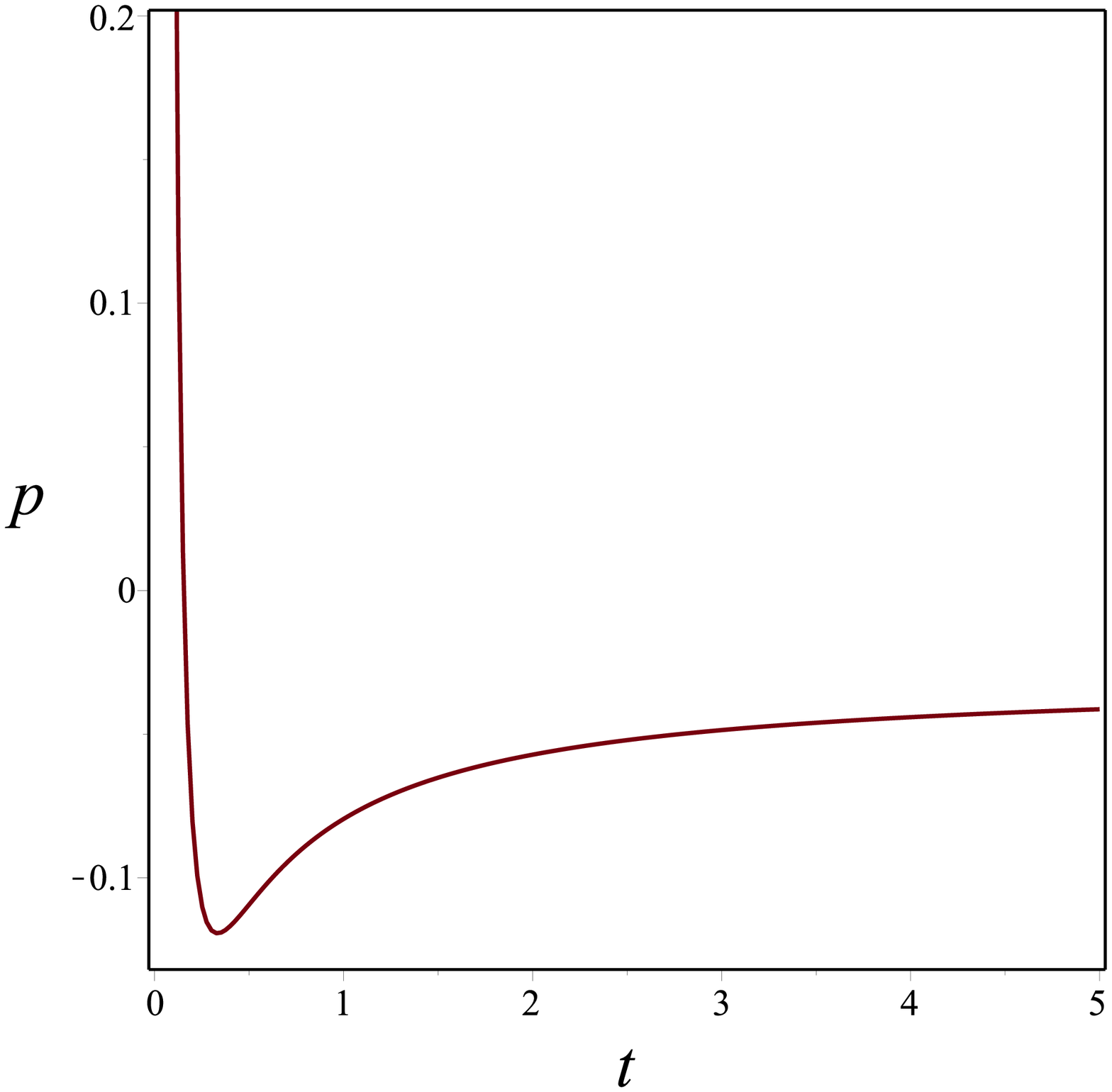}} \\
	\subfigure[$\omega$]{\label{ggg1}\includegraphics[width=0.3\textwidth]{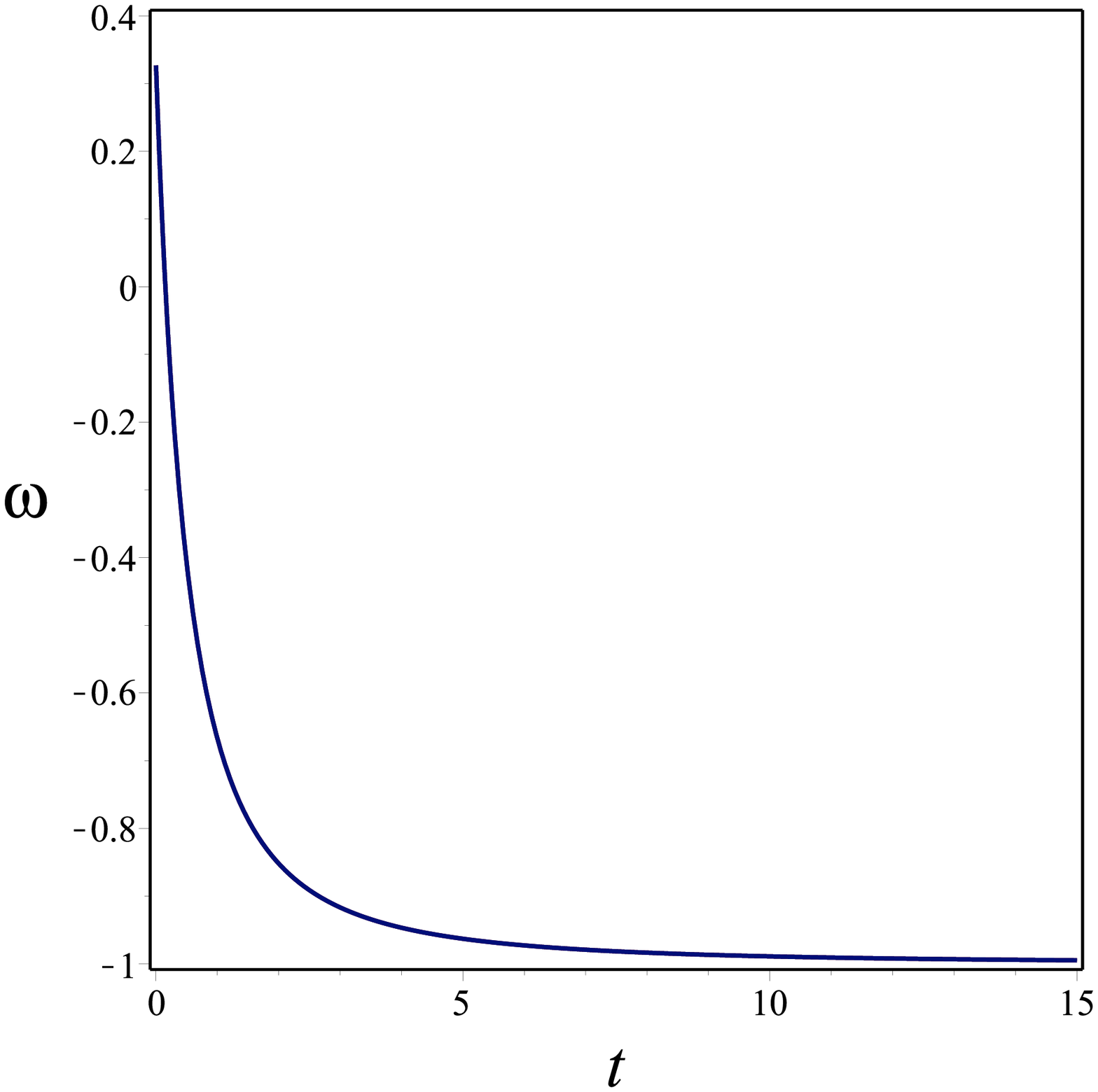}}
	\subfigure[$p$]{\label{gg3}\includegraphics[width=0.32\textwidth]{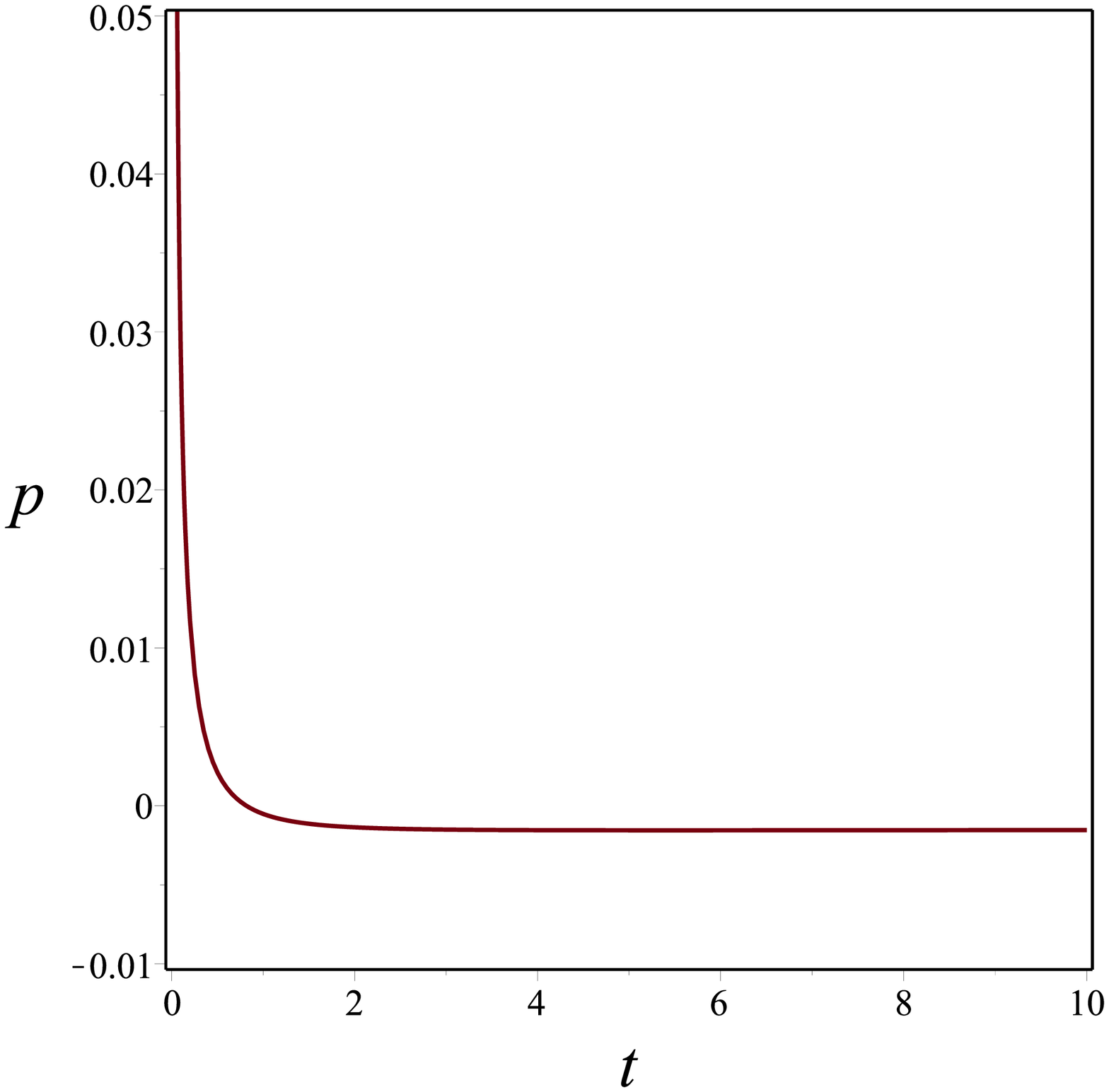}} 
	\subfigure[$\omega$]{\label{ggg3}\includegraphics[width=0.32\textwidth]{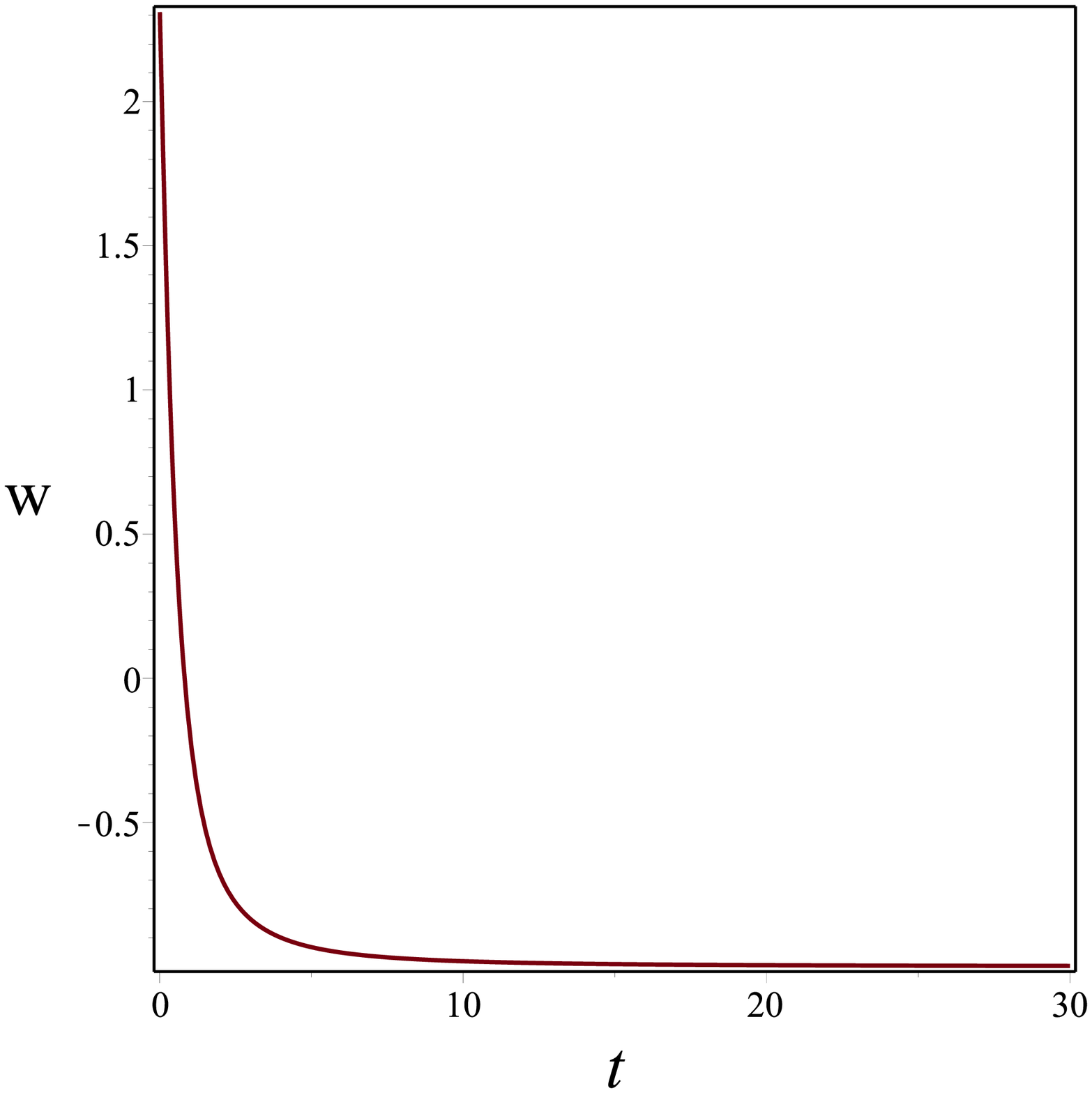}}
	\caption{The behavior of $p(t)$, $\rho(t)$ and $\omega(t)$ in other cosmological frameworks where the same forms of $a(t)$ have been utilized. (a) and (b) show the behavior of $p(t)$ and $\omega(t)$ for the hyperbolic solution in the entropy-corrected cosmology where $-1 \leq \omega(t) \leq \frac{1}{3}$ and the solid lines represent the flat case. (c) and (d) show the behavior of $p(t)$ and $\omega(t)$ for the hybrid solution in the entropy-corrected cosmology where $\omega(t)$ varies in the same range. (e) and (f) show the behavior of $p(t)$ and $\omega(t)$ for the hybrid solution in the Swiss-cheese brane-world cosmology.  The detailed calculations and numerical values adopted in these plots have been given in \cite{ent2}, \cite{n3} and \cite{br1}. }
  \label{nnn}
\end{figure}

In the framework of the entropy-corrected cosmology \cite{basic5}, a stable flat universe has been reached using the hyperbolic ansatz in \cite{ent2} and using the hybrid ansatz in \cite{n3}. In both of these models, the cosmic pressure shows a sign flipping from positive to negative in corresponding to the positive-to-negative sign flipping of the deceleration parameter. Also, there is no an improper behavior in the evolution of the equation of state parameter in both of them as indicated in figure (\ref{nnn}). The sign flipping in the evolution of cosmic pressure has also obtained in the framework of Swiss-cheese brane-worlds by utilizing the hybrid ansatz \cite{br1}. Using the hyperbolic ansatz in universal extra-dimensional cosmology also leads to a sign flipping of cosmic pressure \cite{n1}. It has also been shown that this change of sign in the evolution of cosmic pressure exists in the context of Chern-Simons modified gravity upon using both the hyperbolic and the logamediate inflation forms \cite{n2}. It also appears in the framework of cyclic cosmology \cite{cycl}.

\section{Conclusion}

We have have investigated the cosmic transit and dark energy assumption in a scale-invariance cosmological framework using three different toy models. In order to explain the cosmic transit according to the dark energy assumption, the pressure should be positive in the early-time decelerating epoch and negative in the late-time accelerating epoch. Although cosmic transit exists in the three models, the pressure stays always negative during cosmic evolution. Another bad feature is the evolution of the equation of state parameter which always shows a singularity at specific time. It is interesting to note that a future quintessence-dominated universe with different asymptotic values is predicted in the three models. Finally, We have compared the results we obtained in the framework of the scale-invariance cosmology with other cosmological models in different gravity theories where the same ansatze have been utilized. While the same three forms of $a(t)$ lead to a complete description of $p$ and $\omega$ in other theories, the failure to predict a complete behavior of $p$ and $\omega$ using the same forms in the scale-invariance cosmology puts the cosmological viability of the theory in doubt.

\section*{Acknowledgment}
We are so grateful to the reviewer for his many valuable suggestions and comments that significantly improved the paper.

\end{document}